\documentclass[twocolumn]{aastex631}

\usepackage{xcolor}
\usepackage{graphicx}
\begin{document}

\title{Reconnection generated plasma flows in the quasi-separatrix layer in localised solar corona}

\author{Sripan Mondal}
\affiliation{Department of Physics, Indian Institute of Technology (BHU), Varanasi-221005, India}
\author{A.K. Srivastava}
\affiliation{Department of Physics, Indian Institute of Technology (BHU), Varanasi-221005, India}
\author{Sudheer~K.~Mishra}
\affiliation{Indian Institute of Astrophysics, Kormangala, Bangalore, India.}
\author{K.~Sangal}
\affiliation{Department of Physics, Indian Institute of Technology (BHU), Varanasi-221005, India}
\author{Pradeep Kayshap}
\affiliation{Vellore Institute of Technology (VIT) Bhopal, 466114, India}
\author{Yang~Guo}
\affiliation{School of Astronomy and Space Science, Nanjing University, China}
\author{David~I~Pontin}
\affiliation{School of Information and Physical Sciences, University of Newcastle, Australia}
\author{Vadim M. Uritsky}
\affiliation{Catholic University of America and NASA-GSFC, USA}
\author{Leon~Ofman}
\affiliation{Catholic University of America and NASA-GSFC, USA}
\affiliation{Visiting, Tel Aviv University, Israel}
\author{T.-J.~Wang}
\affiliation{Catholic University of America and NASA-GSFC, USA}
\author{Ding~Yuan}
\affiliation{Institute of Space Science and Applied Technology, Harbin Institute of Technology, Shenzhen 518055, China}



\begin{abstract}


Multiwavelength observations of the propagating disturbances (PDs), discovered by Atmospheric Imaging Assembly (AIA) onboard Solar Dynamics Observatory (SDO), are analyzed to determine its driving mechanism and physical nature. Two magnetic strands in the localised corona are observed to approach and merge with each other followed by the generation of brightening, which further propagates in a cusp-shaped magnetic channel. Differential emission measure analysis shows an occurrence of heating in this region-of-interest (ROI). We extrapolate potential magnetic field lines at coronal heights from observed Helioseismic and Magnetic Imager (HMI) vector magnetogram via Green's function method using MPI-AMRVAC. We analyze the field to locate magnetic nulls and quasi-separatrix layers (QSLs) which are preferential locations for magnetic reconnection. Dominant QSLs including a magnetic null are found to exist and match the geometry followed by PDs, therefore, it provides conclusive evidence of magnetic reconnection. In addition, spectroscopic analysis of Interface Region Imaging Spectrograph (IRIS) Si IV 1393.77 \AA~line profiles show a rise of line-width in the same time range depicting presence of mass motion in the observed cusp-shaped region. PDs are observed to exhibit periodicities of around four minutes. The speeds of PDs measured by Surfing Transform Technique are almost close to each other in four different SDO/AIA bandpasses, i.e., 304, 171, 193 and 131 {\AA} excluding the interpretation of PDs in terms of slow magnetoacoustic waves. We describe comprehensively the observed PDs as quasi-periodic plasma flows generated due to periodic reconnection in vicinity of a coronal magnetic null.      

\end{abstract}


\section{INTRODUCTION} 

Corona, the outermost part of the solar atmosphere, is known to sustain a very high temperature (roughly of the order of Mega Kelvin (``MK") compared to the inner layers. The corona drains out its energy through thermal conduction, solar wind outflow as well as via radiation to the surroundings. Therefore, in order to sustain the high temperature, the corona must be subjected to continuous heating, with possible mechanisms such as by dissipation of {Alfv\'en} waves  
\citep{2011ApJ...736....3V,2017NatSR...743147S,2020SSRv..216..140V} and small-scale energy release events such as nanoflares \citep{1988ApJ...330..474P} as a manifestation of magnetic reconnection. Several energy dissipation mechanisms such as resonant absorption \citep{1978ApJ...226..650I}, phase mixing \citep{1983A&A...117..220H} have been proposed through which wave heating may take place. \par

As per the coarser definition, the magnetic reconnection is a breaking and reconfiguration of oppositely directed magnetic field lines in highly conducting plasma via the formation of extended magnetic singularities in the form of current sheet (i.e., localized region with high current density resulting from steep gradient of the magnetic field over short length scale) either associated with MHD instabilities \citep{2000A&A...353.1074B,2017JPlPh..83e2001V} or triggered by external perturbations \citep{1985PhFl...28.2412H,2019ApJ...887..137S,2021ApJ...920...18S}. Magnetic reconnection is able to release magnetic energy stored in the large-scale magnetic field, hence serving to relieve the stress in a non-potential field. The field lines coming into the current sheet region are referred to as reconnection inflows, while the reconfigured field lines dragging plasma with them as they exit outside the diffusion region are observed to move away as outflows \citep{2007mare.book.....P}. Magnetic reconnection has several effects like-  conversion of magnetic energy into heat via Ohmic dissipation, acceleration of plasma by generating bulk kinetic energy from magnetic energy, generation of shock waves, turbulence, filamentation of currents, etc \citep{2022LRSP...19....1P}. 
Magnetic reconnection is essentially a three-dimensional process which is classically linked to the presence of null points where magnetic field vanishes whenever it is studied in two-dimensional regime \citep{1990ApJ...350..672L,1997SoPh..174..265W}. Actually, the magnetic forces in the neighbourhood of magnetic nulls are not sufficient to withstand large variations of the magnetic stress, therefore, result in the collapse of magnetic topology followed by the generation of current singularities (in ideal plasma conditions) and magnetic reconnection \citep{1996RSPTA.354.2951P}. Although this classical explaination provides fair enough idea about the magnetic reconnection process, still it is not sufficient while looking for reconnection scenarios in the presence of complex magnetic geometries in the real 3D solar atmosphere. \par

For example, in the context of the solar atmosphere, magnetic reconnection may take place along those parts of the photospheric polarity inversion lines where field lines meet the photosphere (known as bald patches) and form current singularities as a result of those attachments \citep{1989A&A...221..287A,1991A&A...243..492V}.  In 3D, the magnetic topology associated with magnetic nulls contains a spine axis and a fan plane. The straight field lines that are directed away or towards the null are collectively termed as the spine. The surface made up of field lines radiating or spiraling around nulls is defined as the fan plane. Depending upon the radial or spiral nature of field lines in the fan plane, these nulls are categorized into radial and spiral types \citep{1988JGR....93.8583G,1996PhPl....3..759P,1997GApFD..84..245P}. Presence of radial nulls has been reported with a few spacecraft observations \citep{2018ApJ...860..128L,2019GeoRL..46.5698C} and therefore confirms their existence in the diffusion regions. On the other hand, spiral nulls are rarely investigated in the diffusion region \citep{2017GeoRL..44...37F}. Also, reconnection can occur along separators which are the intersections of separatrices or separatrix layers (SLs) (i.e., the layers across which magnetic field lines show discontinuity in their connectivity) \citep{1996ASPC..111..331P,2000AdSpR..26..549F,2000SoPh..193....1G}. \citet{1997A&A...325..305D} discussed that all the solar flare events could not be elucidated by these topological features such as magnetic nulls, separators and bald patches. Hence, these classes of topological features need to be extended on the basis of the concept of magnetic field line connectivity.\par
 
Therefore, a more generalised concept has been established that magnetic field lines subject to a drastic change of connectivity form the foundation of current accumulation even if no magnetic nulls are present there \citep{1994ApJ...437..851L}. \citet{1995JGR...10023443P} termed the flux tubes exhibiting such behaviour as quasi-separatrix layers (QSLs). For proper detection of QSLs, squashing factor (Q) has been measured for elemental flux tubes of infinitesimal cross sections that connect opposite polarities at photospheric level \citep{2002JGRA..107.1164T}. As far as observational consequences of presence of QSLs are concerned, plage brightenings and flare kernels in an X1 flare were located at the intersection of QSLs with the photosphere \citep{1998A&A...332..353G}. Moreover, electric currents were found to be concentrated along the boundaries of QSLs. It has been inferred that when the thicknesses of QSLs and current layers associated with them are sufficiently small for reconnection to take place, the magnetic energy stored within QSLs will be released \citep{1997A&A...325..305D,1997SoPh..174..229M}. Therefore, QSLs are treated as preferential regions for the increment of current density and occurrence of magnetic reconnection even in the absence of magnetic nulls and bald patches \citep{1999ApJ...521..889M,2005A&A...444..961A,2008ApJ...675.1614T}.

In recent high resolution extreme ultraviolet (EUV), on-disk observations, the propagating disturbances (PDs), i.e., translational movements of spatially localised intensity enhancements become inevitable observational scenarios in the solar atmosphere \citep{1998ApJ...501L.217D,2007A&A...463..713O,2012A&A...546A..93G}. The spectroscopic signature of plasma flows are also observed using various spectrometers or spectrographs in different magnetic structures in the solar atmosphere \citep{2008ApJ...676L.147H,2011A&A...534A..90D,2014SoPh..289.4501S,2015SoPh..290.2889K,2019ApJ...874...56R}. Apart from gentle plasma flows coupling different layers of the solar atmosphere, some of them may be linked with the impulsive transients or explosive events in the solar atmosphere \citep{2011Sci...331...55D,2019ApJ...873...79C,2020ApJ...894..155S}. To understand the underlying physical mechanism of these PDs or impulsive plasma flows, understanding of their morphological and thermodynamical properties, accurate estimations of propagation speeds and any inherent periodicities, as well as elaboration of magnetic topology or magnetic structure are essential. \par

\citet{1997ESASP.404..571O} and \citet{1998ApJ...501L.217D} reported the first observations of intensity perturbations along solar coronal plumes which they inferred to be slow magnetoacoustic waves. After this discovery, the presence of slow mode waves was reported in closed as well as open loop topologies using data from several instruments, various data analysis procedures and simulation techniques \citep{2000A&A...355L..23D,2001A&A...377..691B,2009A&A...503L..25W,2009ApJ...697.1674M,2012A&A...546A..50K,2012SoPh..279..427K,2013ApJ...779L...7K,2015ApJ...804....4K,2017A&A...600A..37N}. \citet{2012ApJ...754..111O} used 3D MHD model to show the generation of slow magnetoacoustic waves in coronal loops by impulsive flows. \citet{2021ApJ...909..202C} investigated propagating intensity disturbances in five plumes and found that the PDs had higher propagation speeds in hot AIA channels, i.e., 193 and 211 {\AA} than that in the cooler 171 {\AA} channel. The observed speed ratio between 171 and 193 {\AA} channels had an estimated value of 1.3 which was close to the theoretical value (1.25) of slow magnetoacoustic waves. \citet{2018ApJ...868L..33L} observed that disturbances originating from reconnection region and propagating upward across the magnetic dip region of overlying loops with a mean speed of 200 km \(\mathrm{s^{-1}}\) during reconnection process happening between closed coronal loops and overlying open loops. They suggested that these disturbances are essentially the quasi-periodic magnetoacoustic waves. \par 

On the other hand, depending upon observation of time varying blueward asymmetry in spectroscopic Hinode/EUV Imaging Spectrometer (EIS) observations, the observed intensity oscillations were inferred to be caused by quasi-periodic plasma upflows instead of slow mode waves \citep{2010ApJ...722.1013D,2011ApJ...727L..37T}. \citet{2010A&A...510L...2M} interpreted propagating perturbations along polar plumes as quasi-periodically driven high velocity outflows. \citet{2012ApJ...754...43S} inferred that quasi-periodic pulsations (QPPs) propagating along a cusp-shaped loop formed after a flare are mostly episodic outflows rather than slow magnetoacoustic  waves since they have almost same propagation speed irrespective of formation temperatures. \citet{2014ApJ...793...86P} found that the apparent speed of 30-300 km \(\mathrm{s^{-1}}\) is higher in the low temperature channel AIA 171 {\AA} than those in high temperature channels AIA 193 and 211 {\AA} and hence interpreted PDs as plasma outflows. \par 

\citet{2014ApJ...794...79N} explored the moving bright structures along a magnetic loop connecting a pair of negative and positive fields during a coronal bright point (CBP) event having lifetime of around 20 minutes. The average apparent speed of the moving structures were found to be about 380 km \(\mathrm{s^{-1}}\) along with periodicity between 80 and 100 seconds. In addition to the observation, nonlinear force-free field extrapolation has been performed, which showed the possibility of magnetic reconnection taking place during the CBP. Also, they interpreted those moving bright structures as observational outflows after commencement of magnetic reconnection in a CBP. Later, \citet{2016Ap&SS.361..301L} also reported similar kind of event and inferred that the moving structures are observational outflows after the onset of magnetic reconnection. \citet{2009ApJ...705..926B} linked Hinode EIS and XRT observations of AR 10942 with magnetic field modeling. They found that the observed outflows, having speeds of a few to 50 km \(\mathrm{s^{-1}}\), originate in the vicinity of QSLs where magnetic field lines were showing drastic change in the connectivity over a very thin volume. They inferred that magnetic reconnection at QSLs located in between closed field lines of the active region (AR) and large-scale externally connected field lines acting as a driver of the observed active region outflows. \par

In the present paper, we describe the generation and propagation of quasi-periodic plasma flows from an elongated cusp-shaped region, which matches geometrically with QSLs. Detailed multiwavelength imaging (SDO/AIA), Differential Emission Measure (DEM), magnetic field extrapolation and its topology analysis (nulls and QSLs), spectroscopic observations (IRIS Si IV line), wavelet and surfing transform techniques \citep{2013ApJ...778...26U} are extensively utilized to draw a detailed physical picture of merging of two extended, curved bundle of magnetic field lines indicative of magnetic flux-tubes, onset of reconnection in QSLs or SLs (which also includes magnetic null) and generation of quasi-periodic plasma flows. In section 2, the observational scenario and data (SDO/AIA, SDO/HMI, IRIS/SJI, IRIS/Si IV spectra) associated to the present scientific work as well as methods used for preprocessing of them before further analysis are presented. In section 3, the observational (imaging as well as spectroscopic) results and magnetic topology analysis are reported and discussed in a sequential manner. In section 4, we summarize our new scientific findings and draw the conclusions. 

\begin{figure*}
   \hspace{-1.4 cm}
   \includegraphics[height=13 cm,width=1.10\textwidth]{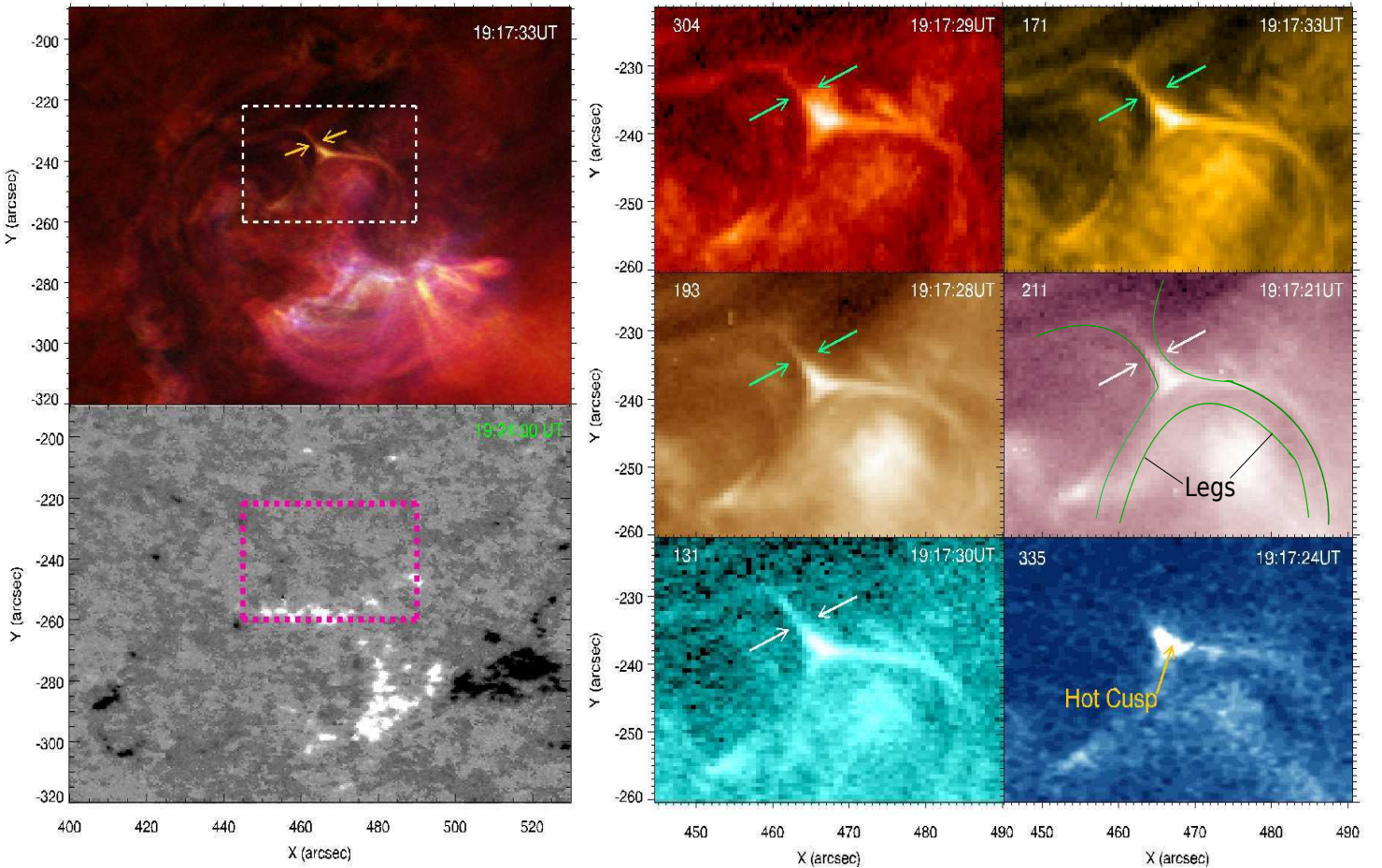}
    \caption{Top-left panel: Composite image of AIA 171, 131 and 193 {\AA} is shown with larger field of view with the white box drawn to demonstrate the more specific FOV of the considered event. Emissions at wavelengths 171, 131 and 193 {\AA} are respectively represented by pseudo-colors golden-yellow, greenish-blue and brown. Bottom-left panel: HMI magnetogram corresponding to the same FOV (magenta box) is shown. Right-panels: Small FOV images in six AIA wavelengths, i.e., 304, 171, 193, 211, 131 and 335 {\AA} are presented. The elongated curved cusp like magnetic structure under consideration is shown by green curves overplotted on 211 {\AA} image. The top part of this cusp-shaped structure is basically the magnetic channel where quasi-periodic plasma flow occured. Although the extended curved field line is not visible properly in 335 {\AA}, the hot cusp-shaped region is clearly evident.}
    \label{label 1}
\end{figure*}

\begin{figure*}
    \includegraphics[height=13 cm,width=18 cm]{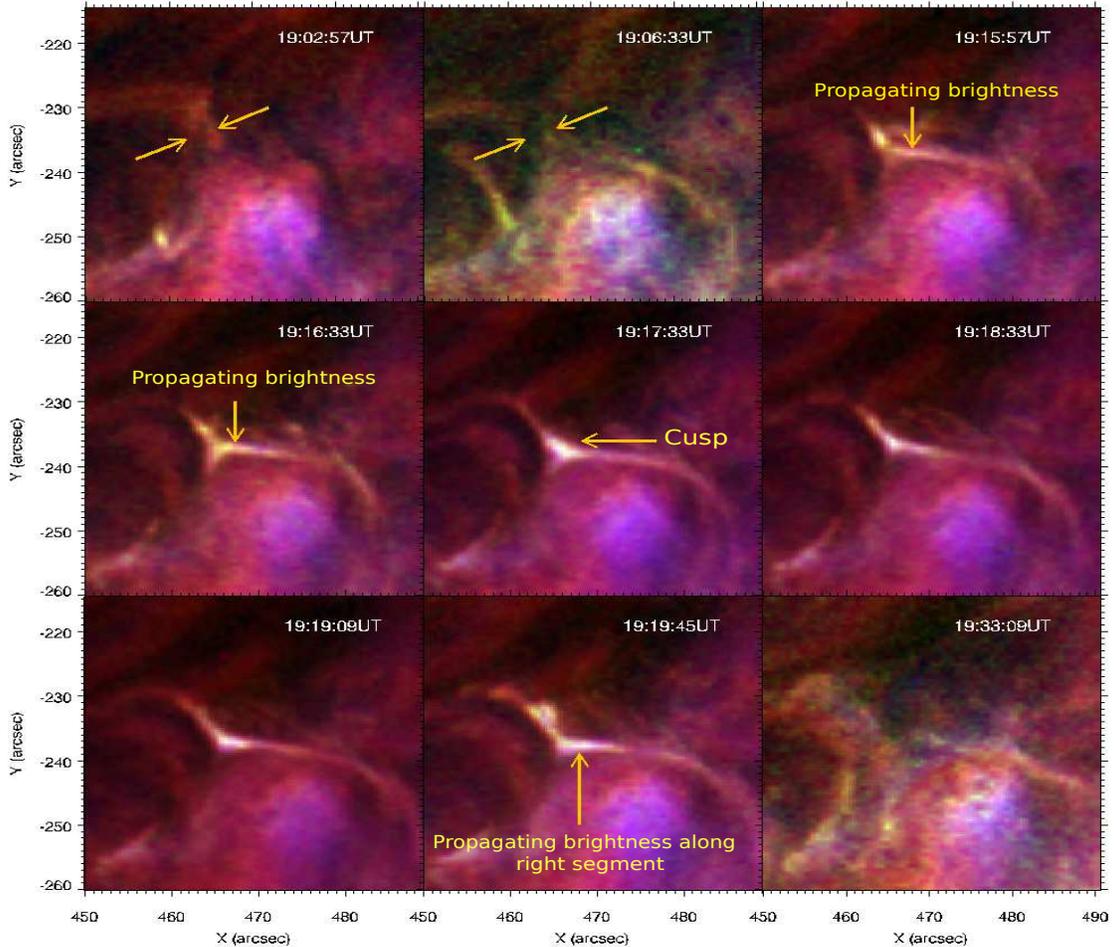}
    \caption{Composite image of 171, 131 and 193 {\AA} show temporal evolution of the event under consideration. Emissions at wavelengths 171, 131 and 193 {\AA} are respectively represented by pseudo-colors golden-yellow, greenish-blue and brown. At first, two curved, elongated magnetic flux-tubes are seen to approach each other around 19:02 UT followed by formation of an apparent current sheet around 19:15 UT. Thereafter, a prominently visible bright Y-shape structure is formed around 19:16 UT. Brightenings are observed to be originated at the cusp-shaped region as shown in rightmost part of middle panel. It propagates downwards and then keeps traversing prominently in right leg of this Y-shape structure. Around, 19:33 UT, the structure gets disappeared. The entire dynamics happening in the time interval 18:48-19:33 UT with temporal cadence being 12 seconds is shown in the animation. The real-time animation consists of the solar event of duration 45 minutes.}
   \label{label 2}
\end{figure*}

\begin{figure*}

\mbox{
\hspace{-0.5 cm}
\includegraphics[scale=0.4,height=8 cm,width=5.5 cm]{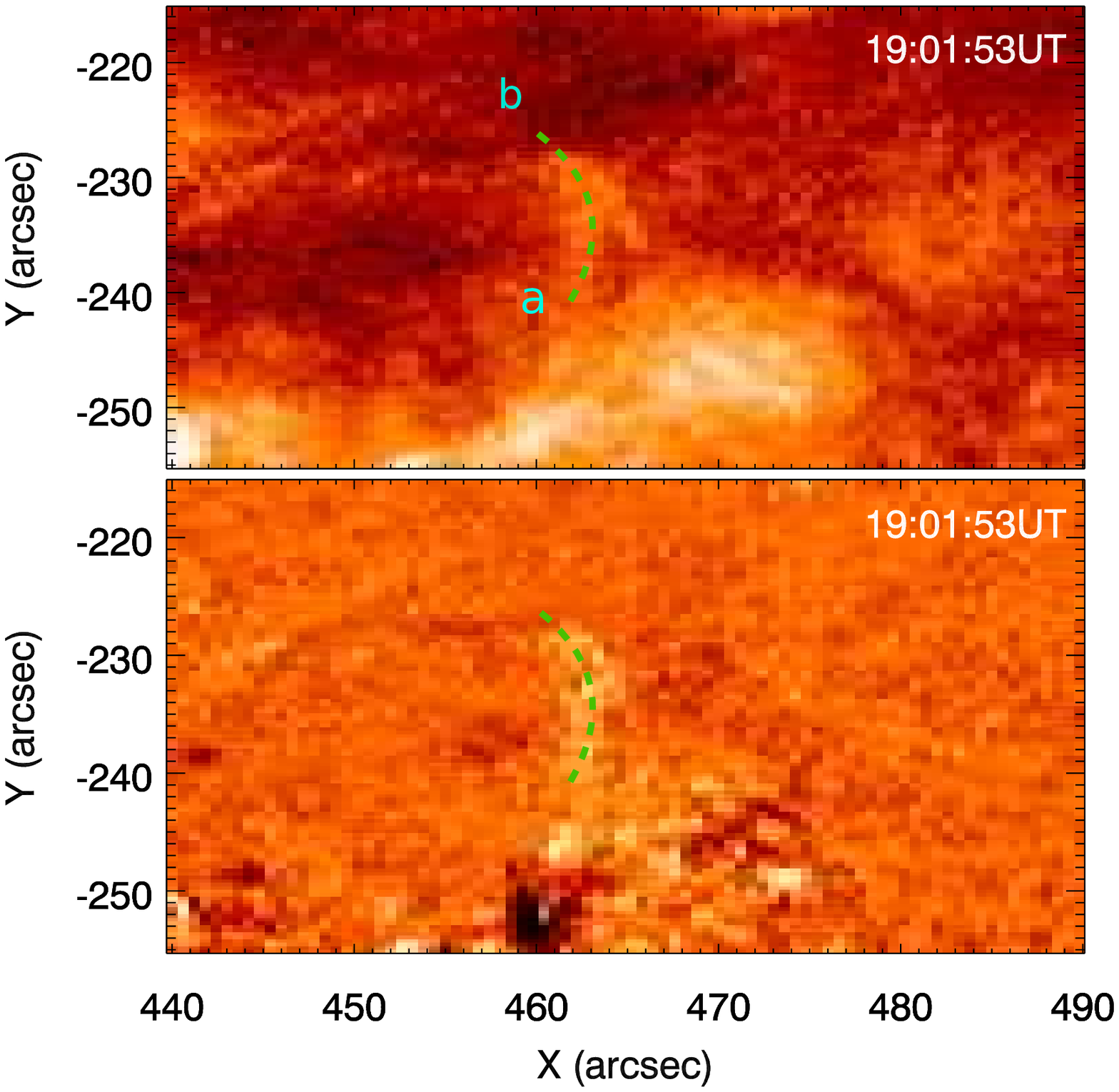}
\includegraphics[scale=0.6,height=8 cm,width=13.5 cm]{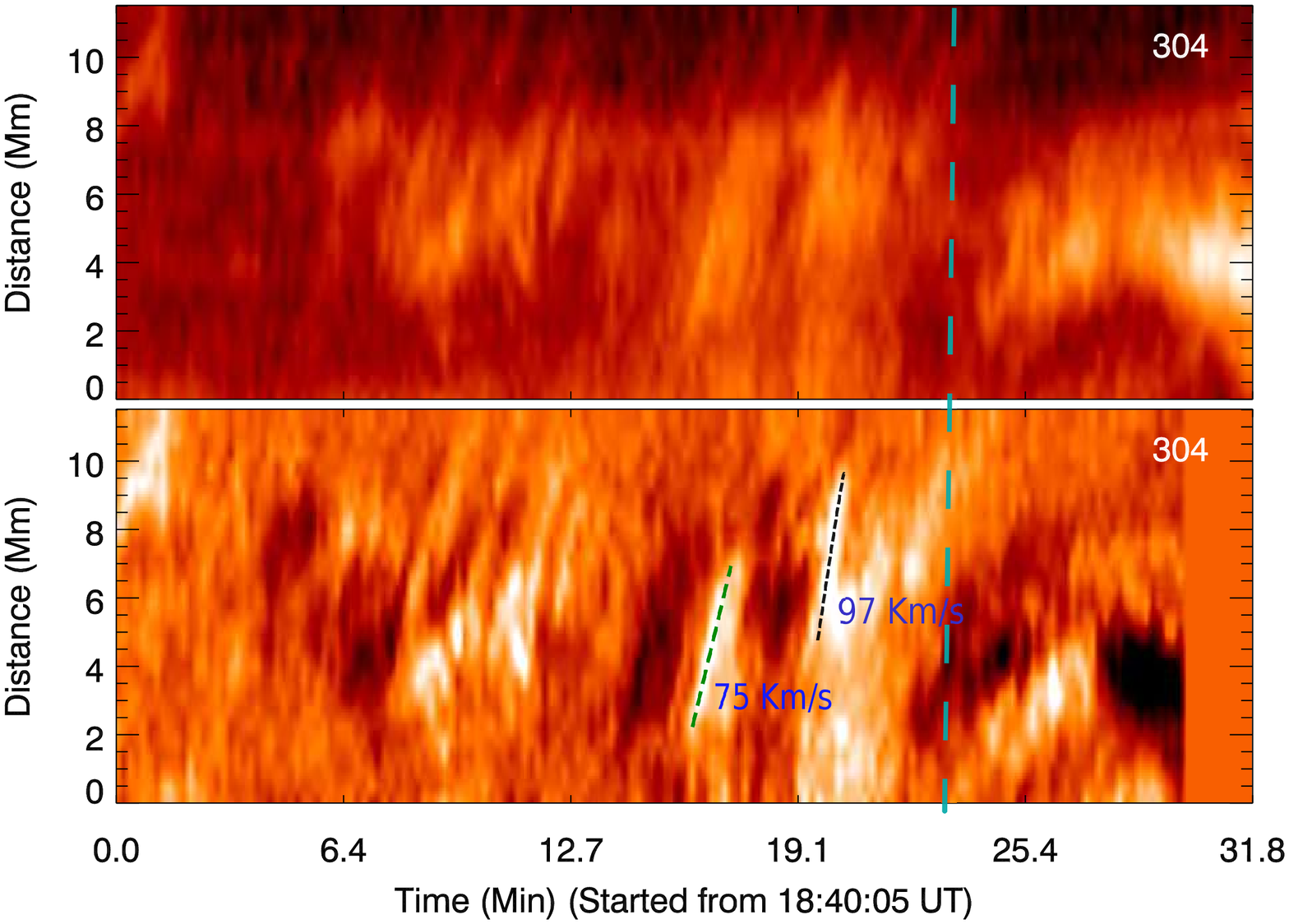}
}
\caption{Left panel: Slit orientation overplotted as green dashed line from a to b on AIA 304 {\AA} image (above: normal intensity; below: running difference with time difference being 2 minutes) having spatial extent X = [440\({\arcsec}\), 490\({\arcsec}\)], Y = [-255\({\arcsec}\), -215\({\arcsec}\)]) for capturing the pre-existing plasma upflows observed to propagate upwards from left magnetic channel to the elongated cusp-shaped region subsequently filling the mass there. Right panel: Distance-time maps (above: normal intensity; below: running difference) provide an evidence of flows with speeds 75 and 97 km \(\mathrm{s^{-1}}\). These speeds were estimated by the slopes of straight line fits along tilted bright fronts in running difference distance-time map. The vertical dashed cyan line denotes the approximate onset time of the merging of flux tubes and hence clearly differentiate this upflow with the reconnection generated plasma flows.}
\label{label 3}
\end{figure*}

\begin{figure*}
\mbox{    
\includegraphics[height=18 cm, width=8 cm,angle=90]{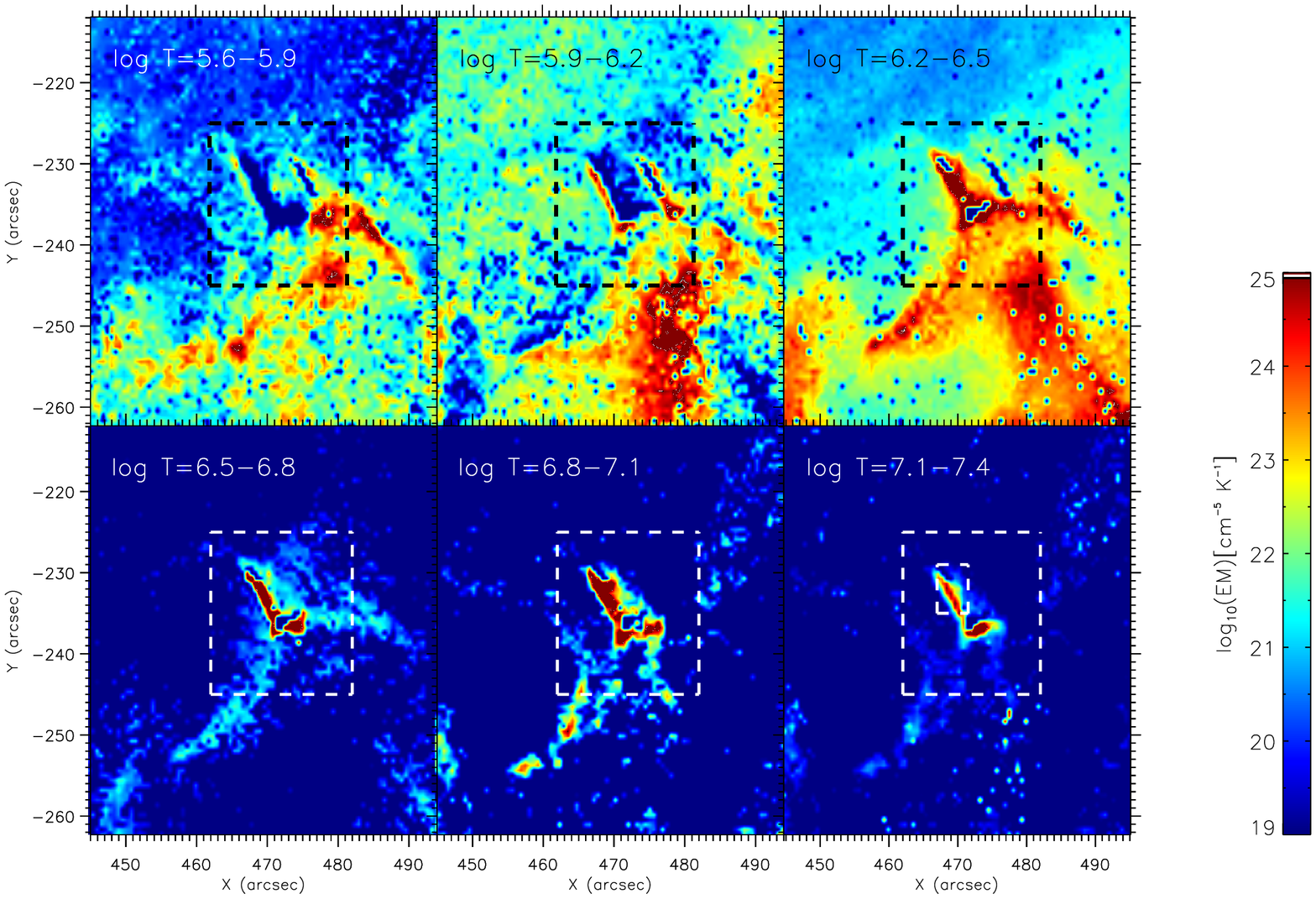}
}
\mbox{
\includegraphics[height=18 cm, width=8 cm,angle=90]{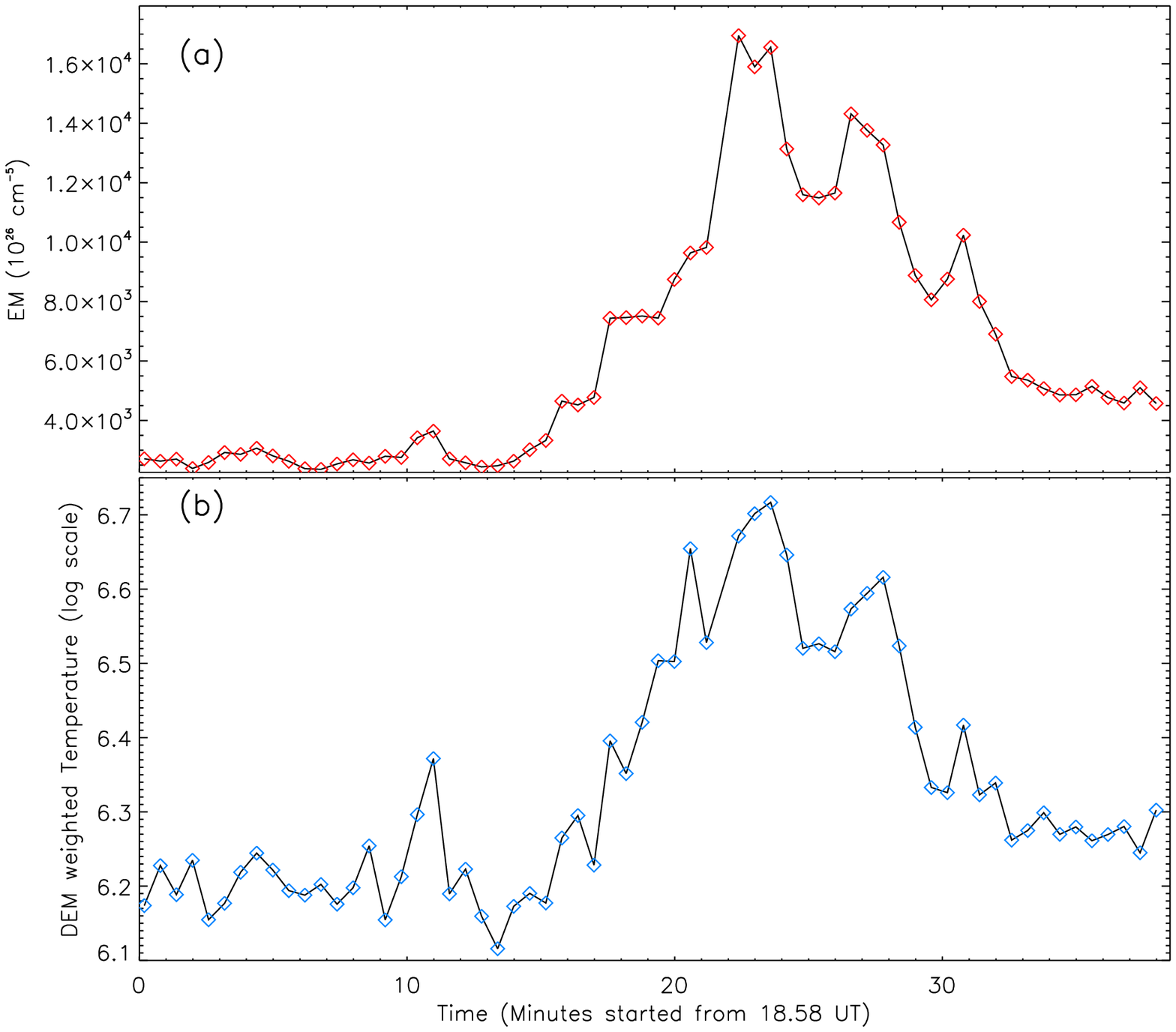}
}
\caption{Top-panel: Differential emission measure maps at six different temperature ranges between log T = 5.6 to 7.4 at time 19:21 UT. Rectangular boxes are used to specify more exact location of the considered event, i.e., brightening in an inverted Y-shape structure. Bottom-panel: (a) Temporal evolution of the total emission measure (b) Evolution of DEM weighted temperature with time covering our considered event evaluated from the region enclosed by smaller rectangular box shown in DEM for log T = 7.1-7.4. The thermal evolution of this structure and its surroundings from 18:50 UT to 19:36 UT with temporal cadence of 36 seconds is shown in the animation containing maps from all the six temperature ranges shown in the top panel of this figure. The real-time animation consists of the event of duration 46 minutes.}
\label{label 4}
\end{figure*}

\begin{figure*}
\mbox{
\hspace{1.0 cm}
\includegraphics[scale=0.8,height=8 cm, width=15 cm]{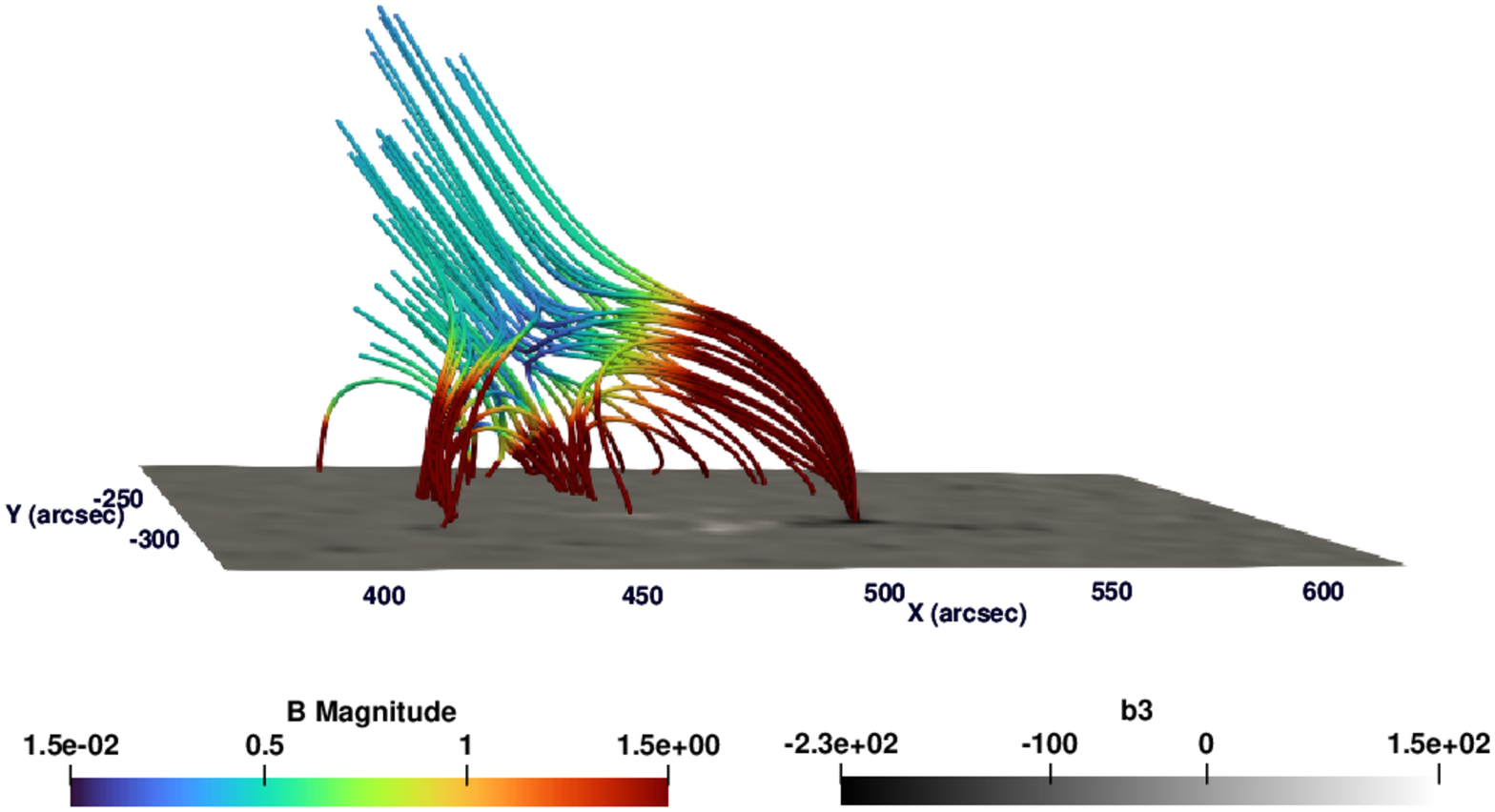}
}
\mbox{
\hspace{1.0 cm}
\includegraphics[scale=0.8, height=8 cm,width=16 cm]{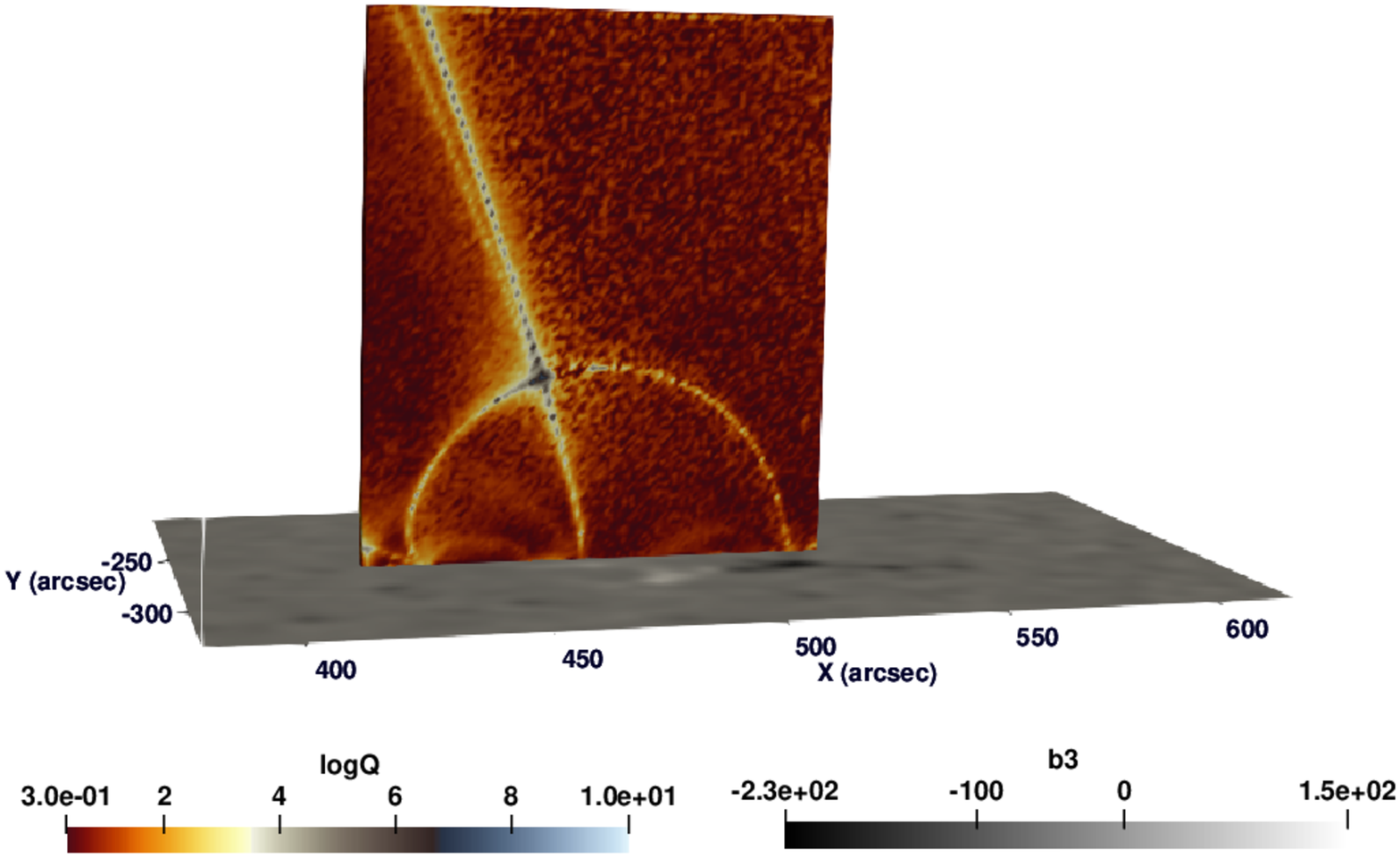}
}

\caption{Top-panel: Extrapolated field lines in the ROI due to potential field extrapolation using MPI-AMRVAC. Bottom-panel: Surface representation of Q (squashing factor) distribution in the ROI resembling the extrapolated magnetic field structures. In both the Figs, b3 stands for line-of-sight (LOS) component of vector magnetic field observed by SDO/HMI. b3 and B-magnitude are shown in the dimensionless code units used in MPI-AMRVAC. To convert these two quantities to units of Gauss, the values shown in respective colorbars should be multiplied by 2. In addition, the values shown in colorbar of B-magnitude are manually rescaled from exact values for the better visualization. Also it is to be noted that the $x$ and $y$ coordinates are shown in helioprojective cartesian coordinate system to match with FOV of AIA observations.}
\label{label 5}

\end{figure*}

\begin{figure*}
\mbox{
\hspace{1 cm}
\includegraphics[height=6 cm, width =8 cm]{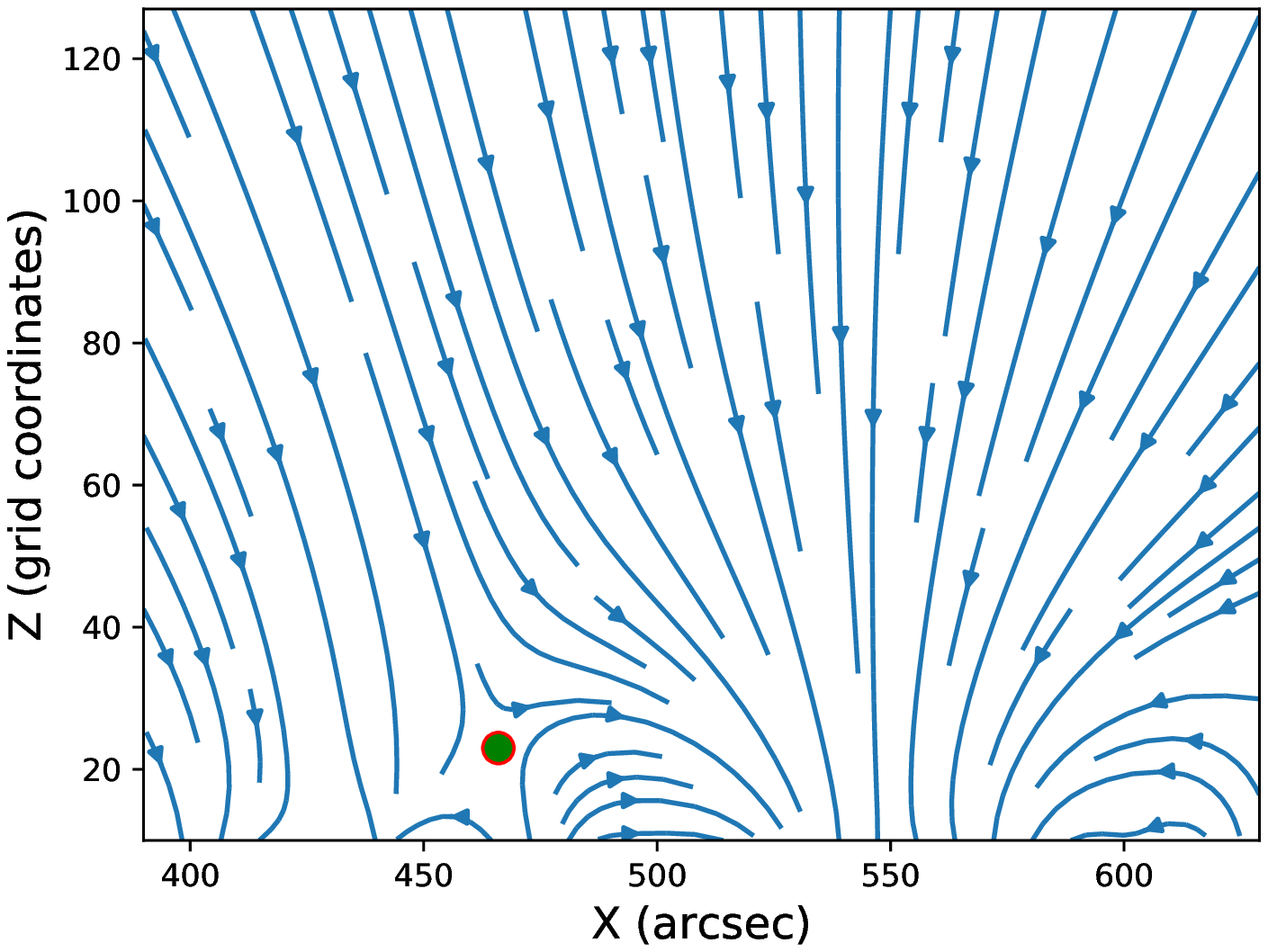}
\includegraphics[height=6 cm, width =8 cm]{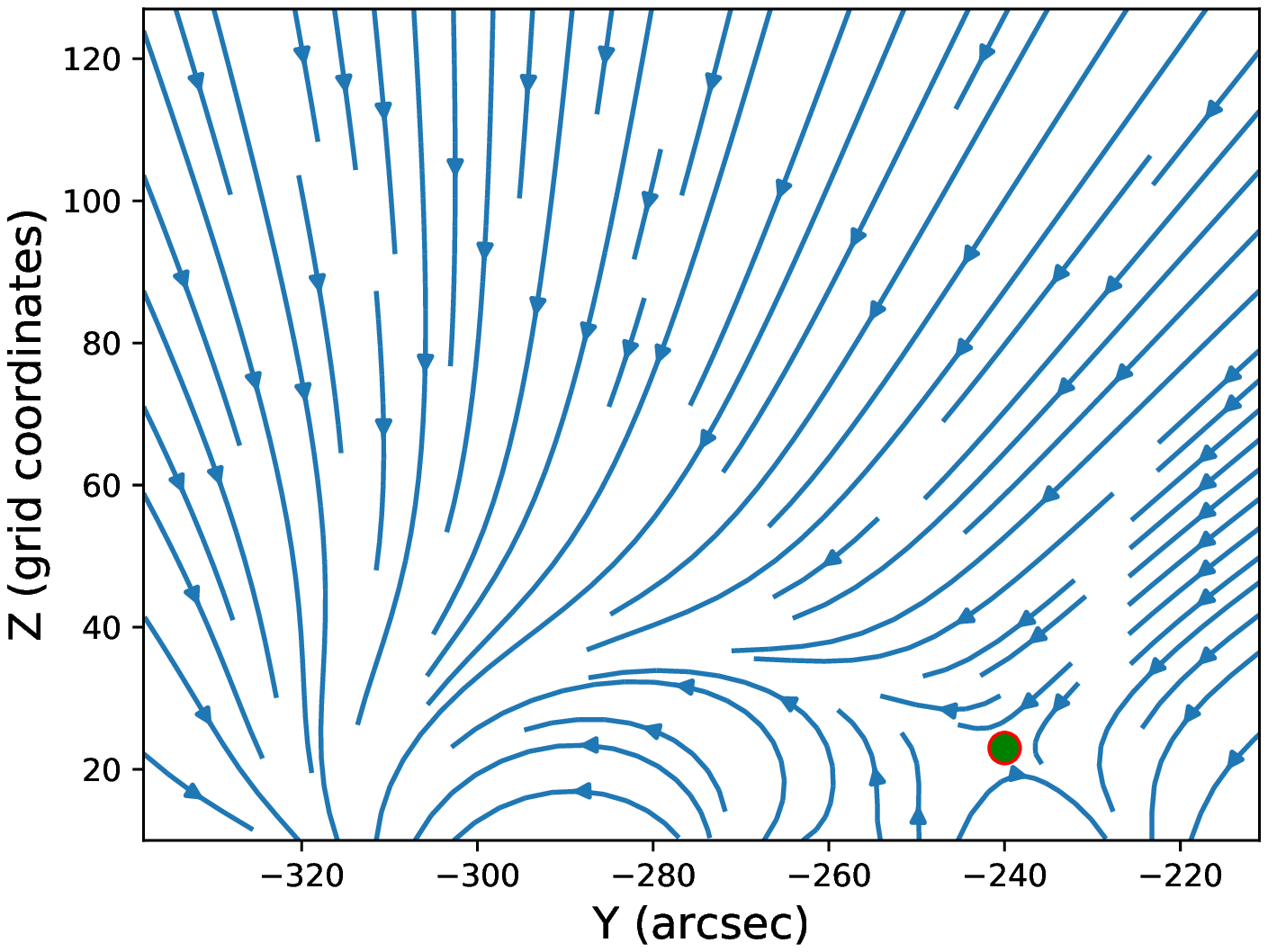}
}
\caption{Left-panel: Magnetic field streamlines and location of magnetic null, shown as green sphere, are displayed while looking along Y-direction. Right-panel: Magnetic field streamlines and location of magnetic null, shown as green sphere, are displayed while looking against X-direction. Spatial resolution along X direction is 1.46 Mm and that along Y and Z direction is 1.40 Mm. The null point (green sphere) found using trilinear method (as shown in the right and left panels here) lies in the bottom part of the spine-line at an exact location of x = 468 \({\arcsec}\) and y = -240 \({\arcsec}\).  The $x$ and $y$ coordinates are shown in helioprojective cartesian coordinate system to match with FOV of AIA observations. } 
\label{label 6}

\end{figure*}

\section{OBSERVATIONAL DATA {\&} ITS OVERVIEW} 

SDO/AIA, with its multiwavelength imaging capability of the full solar disk, investigates various dynamics (e.g. reconnection, outflows, inflows, etc.) happening over small to large spatial length scales in different atmospheric layers. It has a spatial resolution of 1.5 arcseconds and temporal cadence of 12 seconds in all of its EUV filters: 94 {\AA} (Fe X, Fe XVIII, \(T \approx\) 1 MK, \(T \approx\) 6.3 MK), 131 {\AA} (Fe VIII, Fe XXI, Fe XXIII, \(T\approx\) 0,4, 10, 16 MK), 171 {\AA} (Fe IX, \(T\approx\) 0.7 MK), 193 {\AA} (Fe XII, Fe XXIV, \(T\approx\) 1.2 MK, 20 MK), 211 {\AA} (Fe XIV, \(T\approx\) 2 MK), 304 {\AA} (He II, \(T\approx\) 0.05 MK) and 335 {\AA} (Fe XVI, \(T\approx\) 2.5 MK) \citep{2012SoPh..275...17L}. The data analyzed here are observed by AIA within the spatial extent of X = [420\({\arcsec}\), 500\({\arcsec}\)] and Y = [-280\({\arcsec}\), -180\({\arcsec}\)] from 18:40 UT to 19:45 UT on 17 April 2021. We exported level 1.0 data for all EUV AIA wavelengths from Joint Science Operations Center (JSOC)\footnote{\url{http://jsoc.stanford.edu/}} which were already preprocessed to make corrections such as removal of dark current, de-spiking, flat-fielding and bad-pixel removal. Then using informations such as \texttt{CRPIXi}, \texttt{CDELTi} (for i = 1 or 2) etc from FITS headers of AIA files, those level 1.0 data were again processed to co-align all images in different wavelengths to a common centre, to correct the roll angles, and to rescale images to a common plate scale using the Solarsoft \citep{1998SoPh..182..497F} IDL routine \texttt{`aia\_prep.pro'}. Likewise the Helioseismic and Magnetic Imager (HMI), having a spatial resolution of 1.0 arcsecond, provides maps of line of sight component of magnetic field with temporal cadence of 45 seconds, and maps of vector magnetic field with temporal cadence of 720 seconds of the entire disk at the photospheric level \citep{2012SoPh..275..207S}. We used \texttt{`hmi\_prep.pro'} to co-align HMI and AIA data. \par 

We also used spectral data from NASA\textsc{\char13}s Interface Region Imaging Spectrograph (IRIS) spacecraft \citep{2014SoPh..289.2733D}, which observes the lower solar atmosphere (chromosphere and transition region). The IRIS spacecraft observes the Sun in far ultraviolet (FUV: 1331.56--1358.40 {\AA} and 1390.00--1406.79 {\AA}) and near ultraviolet (NUV: 2782.56--2833.89 {\AA}) wavebands using slit jaw imagers (SJI). The IRIS spacecraft also provides spectroscopic observations for some notable bright lines in NUV region, e.g., C II 1334.53 {\AA}, C II 1335.66 {\AA}, Si IV 1393.77 {\AA}, and Si IV 1402.77 {\AA} and in FUV such as Mg II k 2796 {\AA}, Mg II h 2803 {\AA}. The IRIS observation occurred between 17:00--23:00 UT on 17 April 2021. However, we used only Si IV 1393.77 {\AA} spectral data in a short time interval of 19:13--19:30 UT where the reconnection appears to take place. A large coarse 8-step raster scan with a temporal cadence of 5 seconds is used in the analysis, therefore, each raster scan takes 5.0$\times$8 = 40 seconds. The IRIS field of view (FOV) is 120\({\arcsec}\)$\times$119\({\arcsec}\) centered at 449\({\arcsec}\) in the x-direction and -266\({\arcsec}\) in the y-direction. \par

\begin{figure*}
    \hspace{-1.90 cm}
    \includegraphics[height= 20cm,width=0.9\textwidth,angle=90]{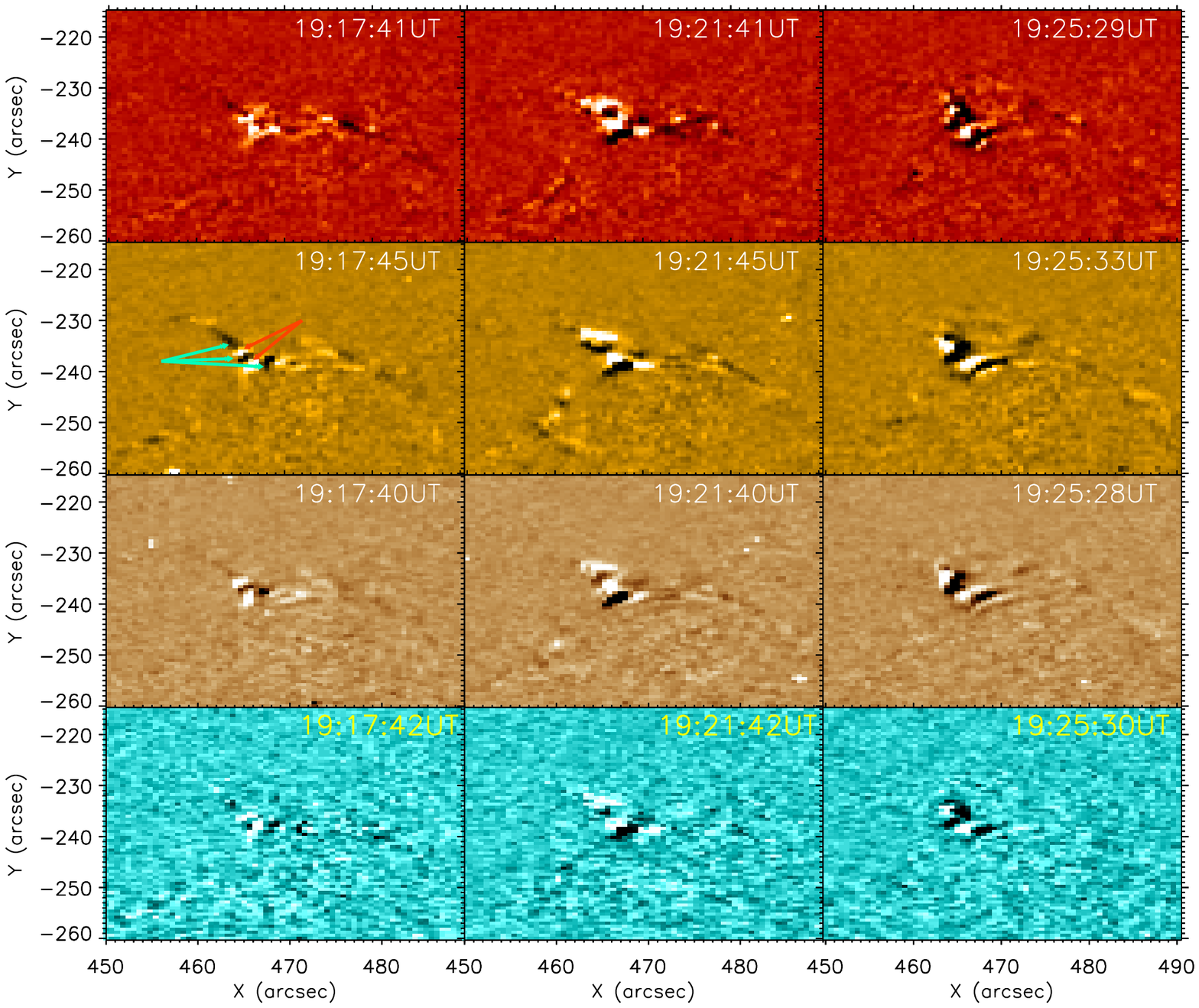}
    \caption{Running difference images at wavelengths 304 {\AA}, 171 {\AA}, 193 {\AA} and 131 {\AA} (successively from top to bottom panel) for time difference of 12 seconds covering different times within our considered temporal range showing successive bright and dark wavefronts propagating downward from the top of the elongated cusp-shaped region. Cyan arrows overplotted on left panel of 171 images are showing dark fronts whereas red arrows are showing bright ones. The full dynamics happening  during the time duration from 19:03:21 UT to 19:32:57 UT with temporal cadence being 12 seconds is shown in the animation (only for 171 {\AA} exhibited in the middle panel). The real-time animation consists of the solar dynamical plasma process of the duration of around 30 minutes.}
    \label{label 7}
\end{figure*}

\begin{figure*}
	\centering
	\includegraphics[scale=0.6,height=10 cm,width=18 cm]{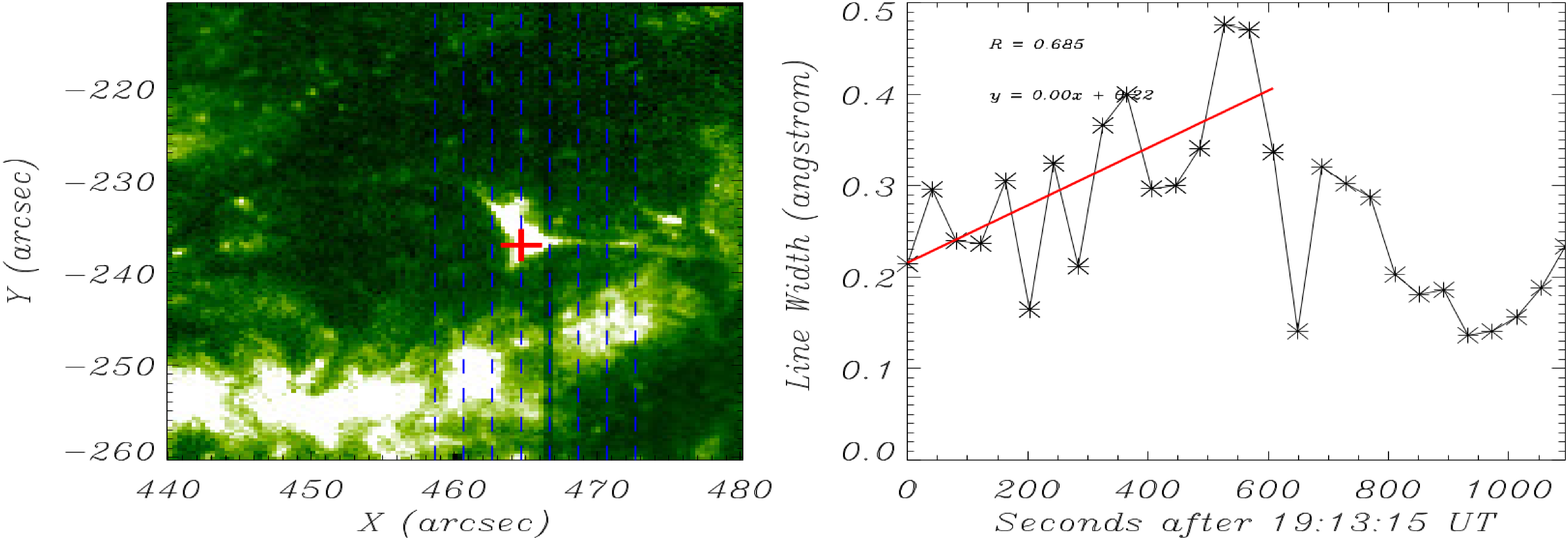}

	\caption{Left-panel: Eight slit positions corresponding to 8-step IRIS observation are shown as blue dashed lines overplotted on IRIS/SJI C-II 1330 {\AA} image. Red cross navigates more specific region from where Si IV line spectra have been taken. Right-panel: Temporal evolution of line width in 1393.77 {\AA} of Si IV line in that region. Red-line fit indicates increasing trend of line-width during the development of the reconnection event.}
	\label{label 8}

\end{figure*}

The evolution of two curved, elongated features indicative of magnetic flux-tubes approaching each other and therefore merging to generate propagating brightness is primarily evident in AIA temporal images (Figs. \ref{label 1}-\ref{label 2}). The evolution of plasma in the surroundings, e.g., increment in DEM, i.e., electron number density at high temperature, generation and propagation of brightness are observed and studied in the considered event using temporal image data in various AIA EUV filters as discussed in section 3 in detail. Composite image of AIA 171, 131 and 193 {\AA} around 19:18 UT (top-left panel) along with images of small field of view (right-panels) showing brightened field lines (elongated inverted Y-shaped topology) in 304, 171, 193, 211 and 131 {\AA}, is shown in Fig. \ref{label 1}. To make the elongated inverted Y-shaped topology more clearly evident, the structure is demonstrated by green lines overplotted on the image of 211 {\AA}. Also, images in 335 {\AA} as well as 94 {\AA} (not shown here) contain clear signature of hot cusp-shaped region, however, extended magnetic legs are not properly visible in these images. HMI magnetogram at time 19:24 UT is also shown with magenta rectangular box enclosing the area of interest. It is evident that region of interest (ROI) is just above a plage region having strong opposite polarities associated with it. \par


\begin{figure*}
\mbox{
\includegraphics[scale=0.2,height=9 cm,width=6 cm]{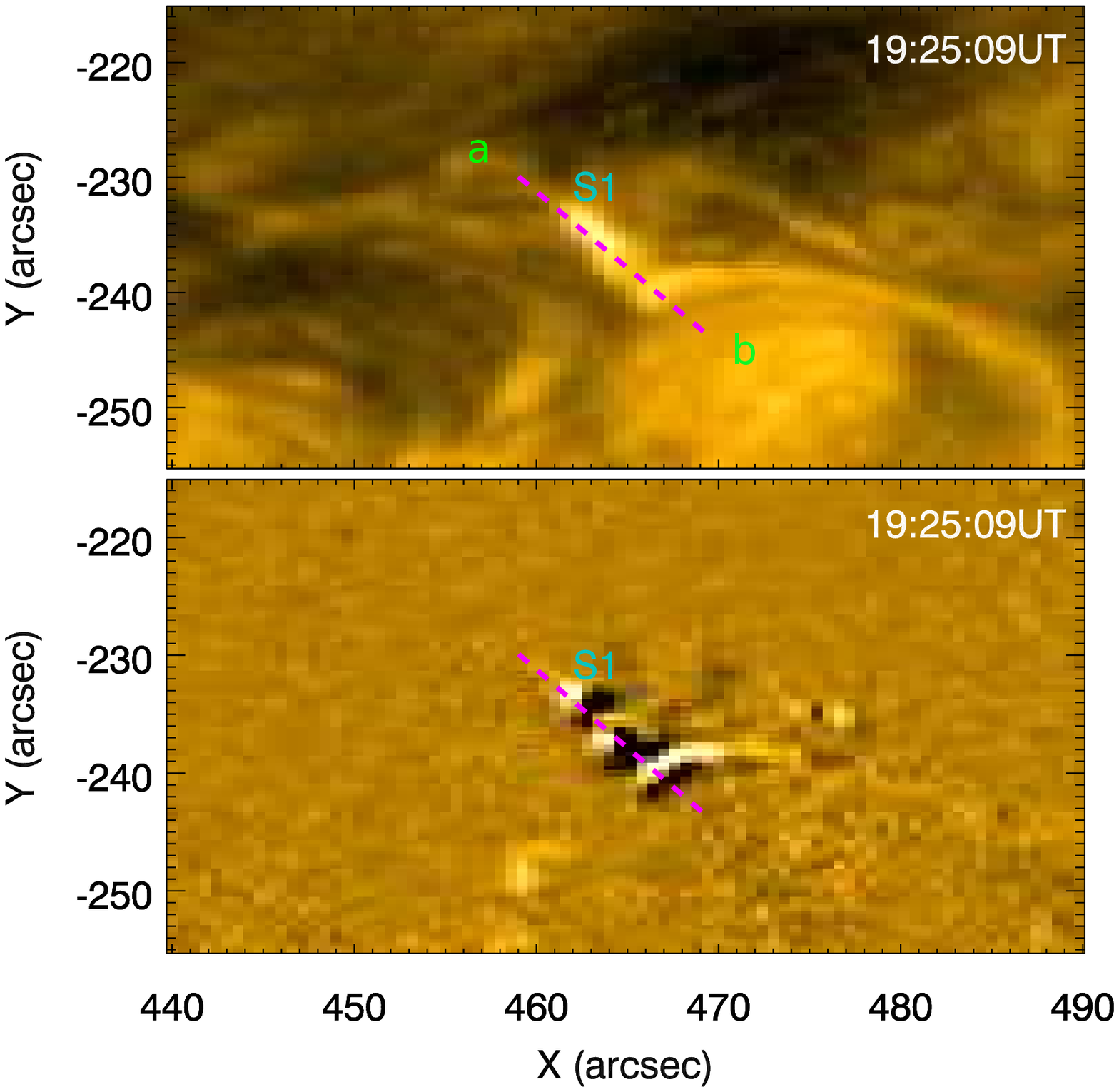}

\includegraphics[scale=0.5,height=13 cm,width=10 cm,angle=90]{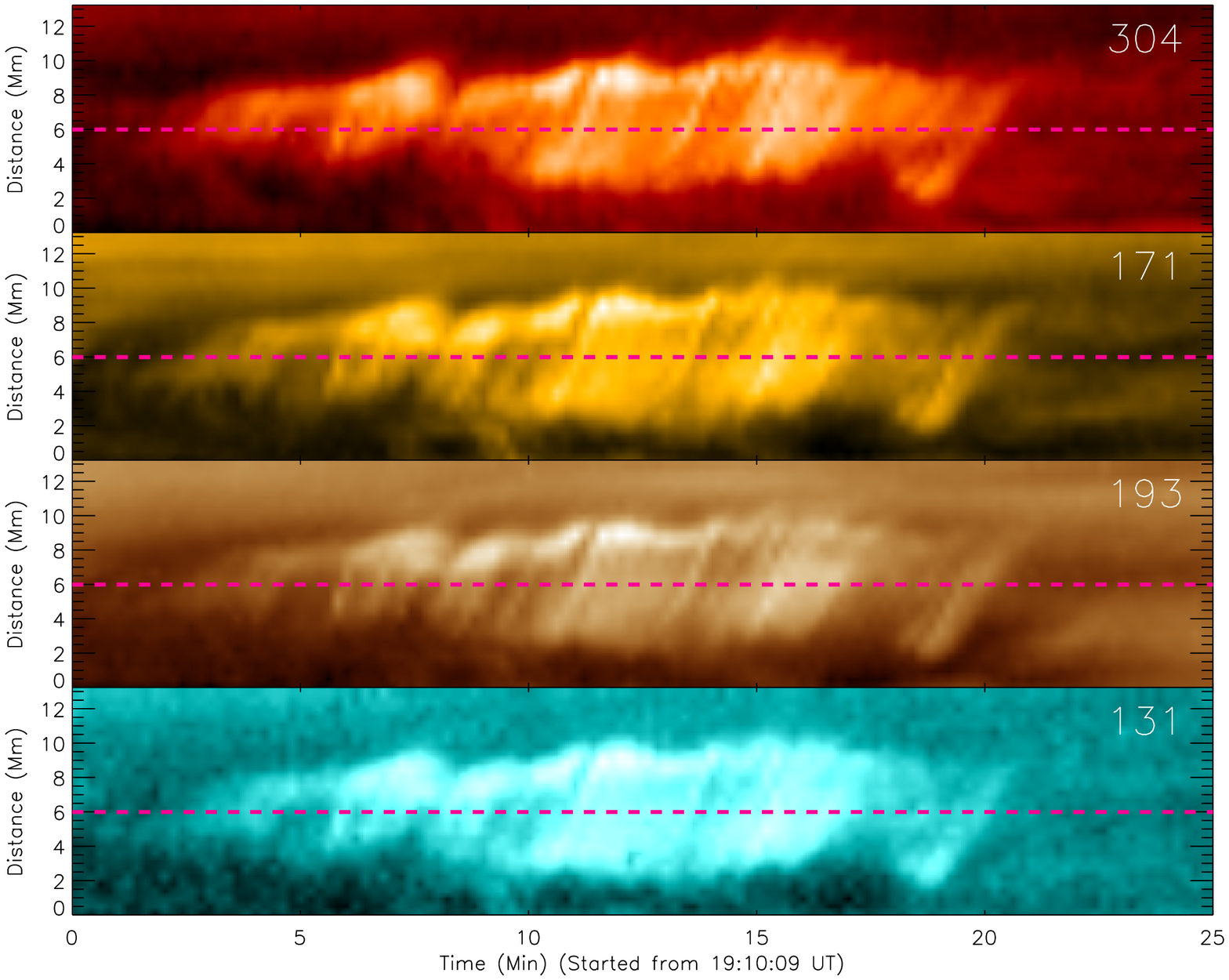}
}
\mbox{
\hspace{0.5 cm}
\includegraphics[scale=0.3,height=20 cm,width=11 cm,angle=90]{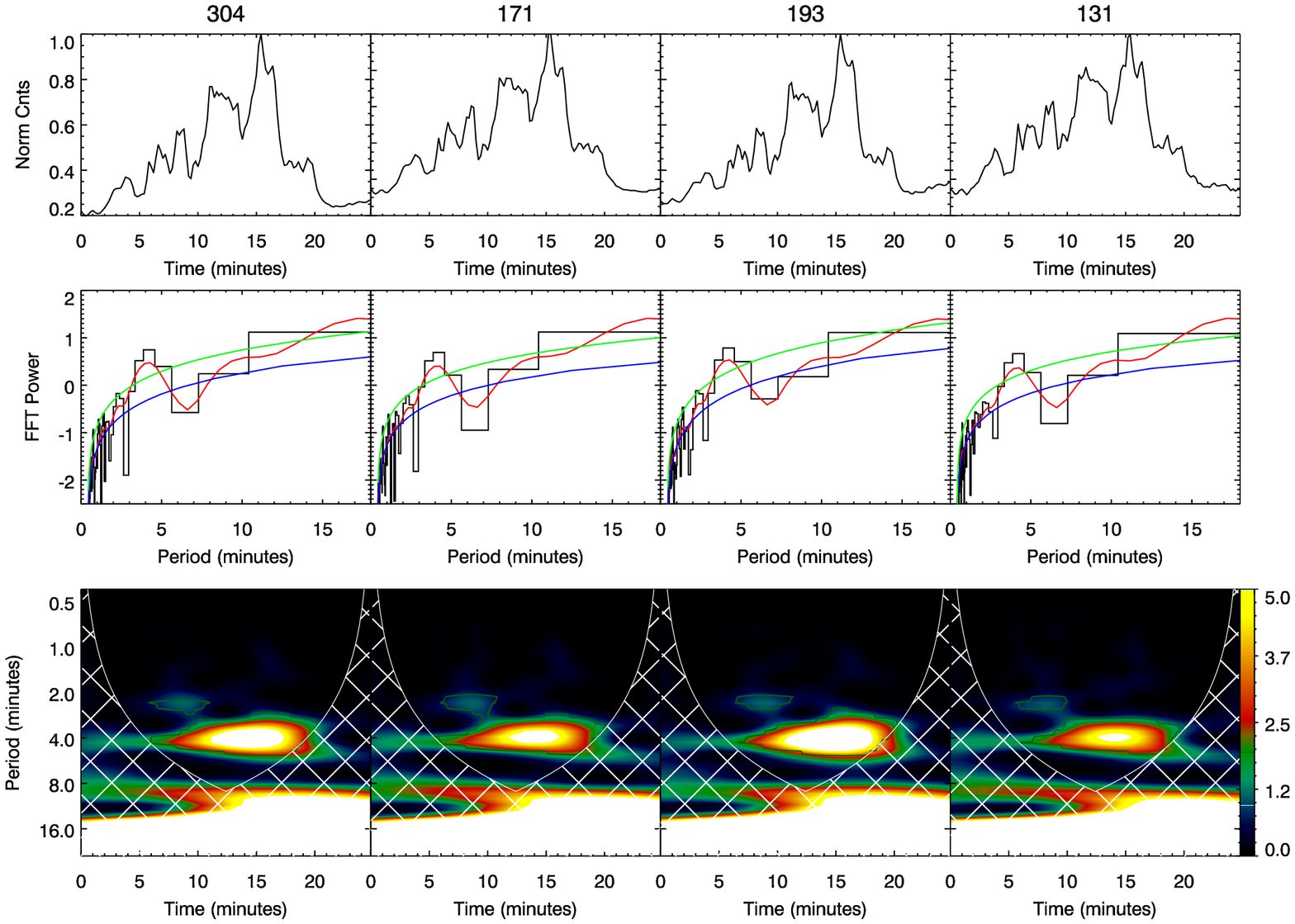}
}
\caption{Top-left panel: Orientation of slit S1 overplotted as magenta dashed line from a to b on AIA 171 {\AA} image (above: normal intensity; below: running difference of 12 seconds) having FOV X = [440\({\arcsec}\), 490\({\arcsec}\)], Y = [-255 \({\arcsec}\), -215 \({\arcsec}\)]) covering the path along which PDs were found to propagate. Top-right panel: Derived distance-time maps at four different wavelengths, i.e., at 304, 171, 193 and 131 {\AA}. Horizontal magenta lines are showing the height (6 Mm) at which the intensity time series were extracted for measurement of their periods. Bottom-panel: Extracted light curves after normalization (top); FFT of the light curves (black color) with global wavelet spectrum (red color), fitted curve using power law (blue color) and 95 {\%} significance level (green color) overplotted on it (middle); Intensity wavelets with contour of significant power constrained by power law are shown for all four chosen wavelengths (bottom).}
\label{label 9}
\end{figure*}

\begin{figure*}
\mbox{
\hspace{-1.0 cm}
\includegraphics[scale=0.4,height=18 cm,width=5.5 cm,angle=90]{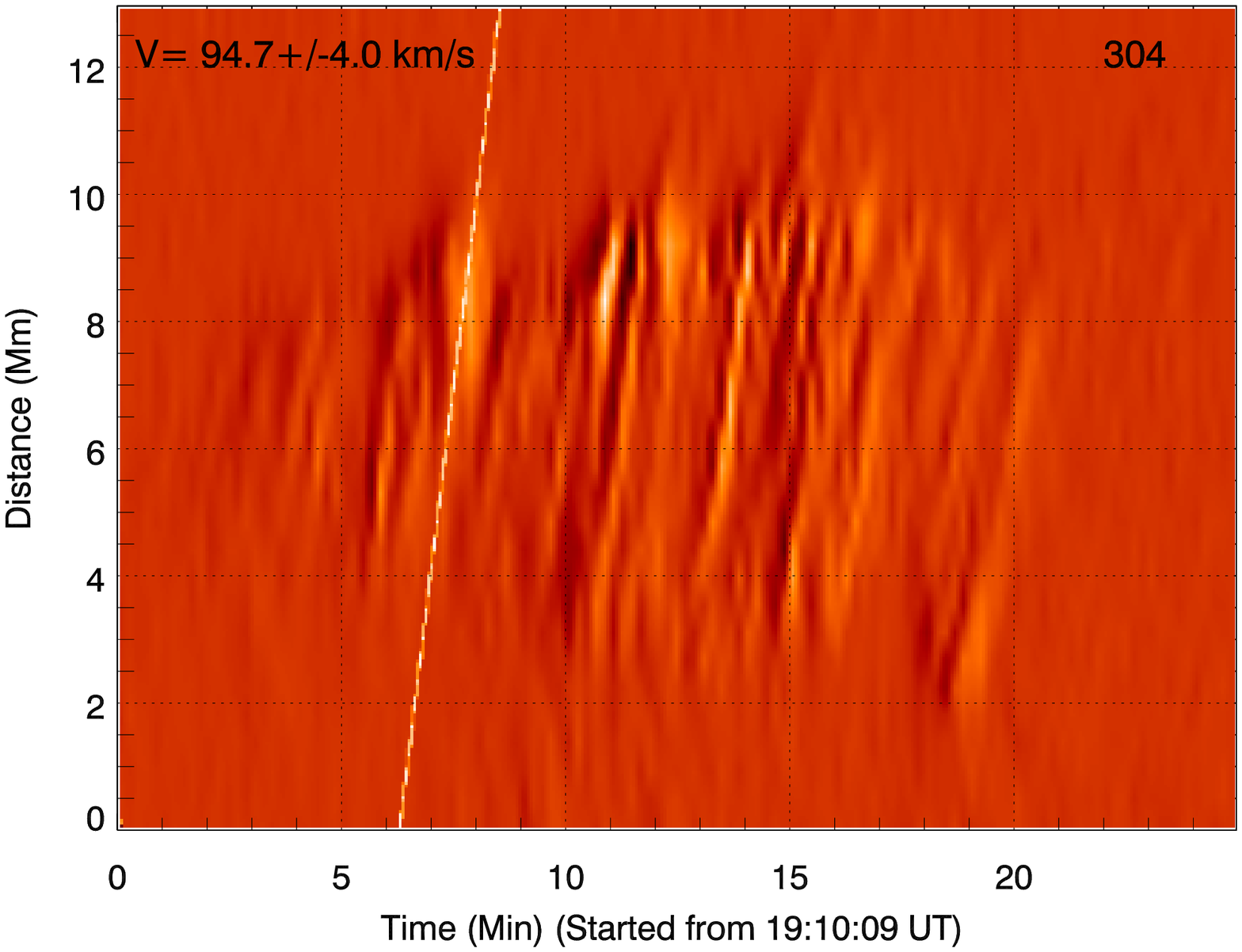}
}
\mbox{
\hspace{-1.0 cm}
\includegraphics[scale=0.4,height=18 cm,width=5.5 cm,angle=90]{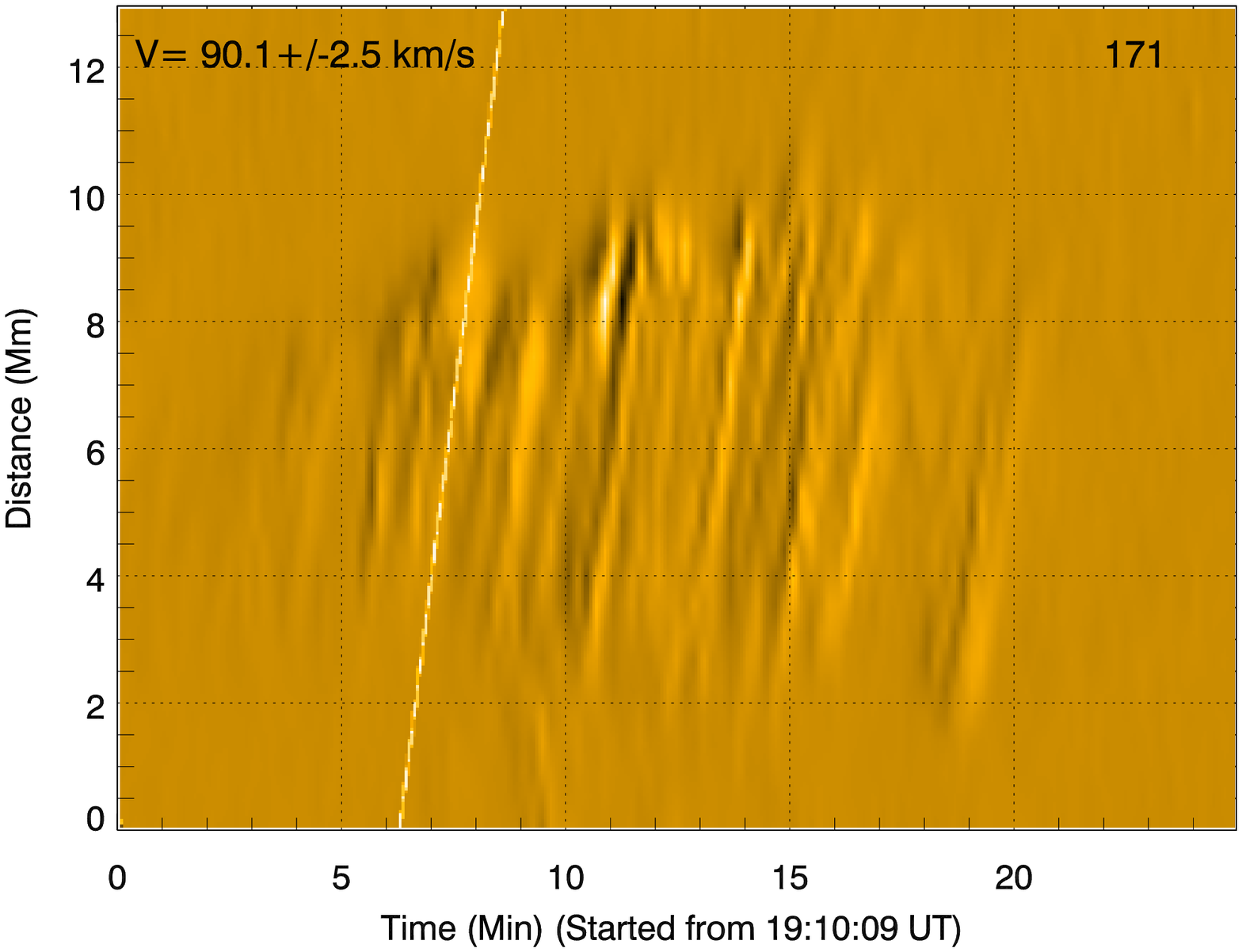}
}
\mbox{
\hspace{-1.0 cm}
\includegraphics[scale=0.4,height=18 cm,width=5.5 cm,angle=90]{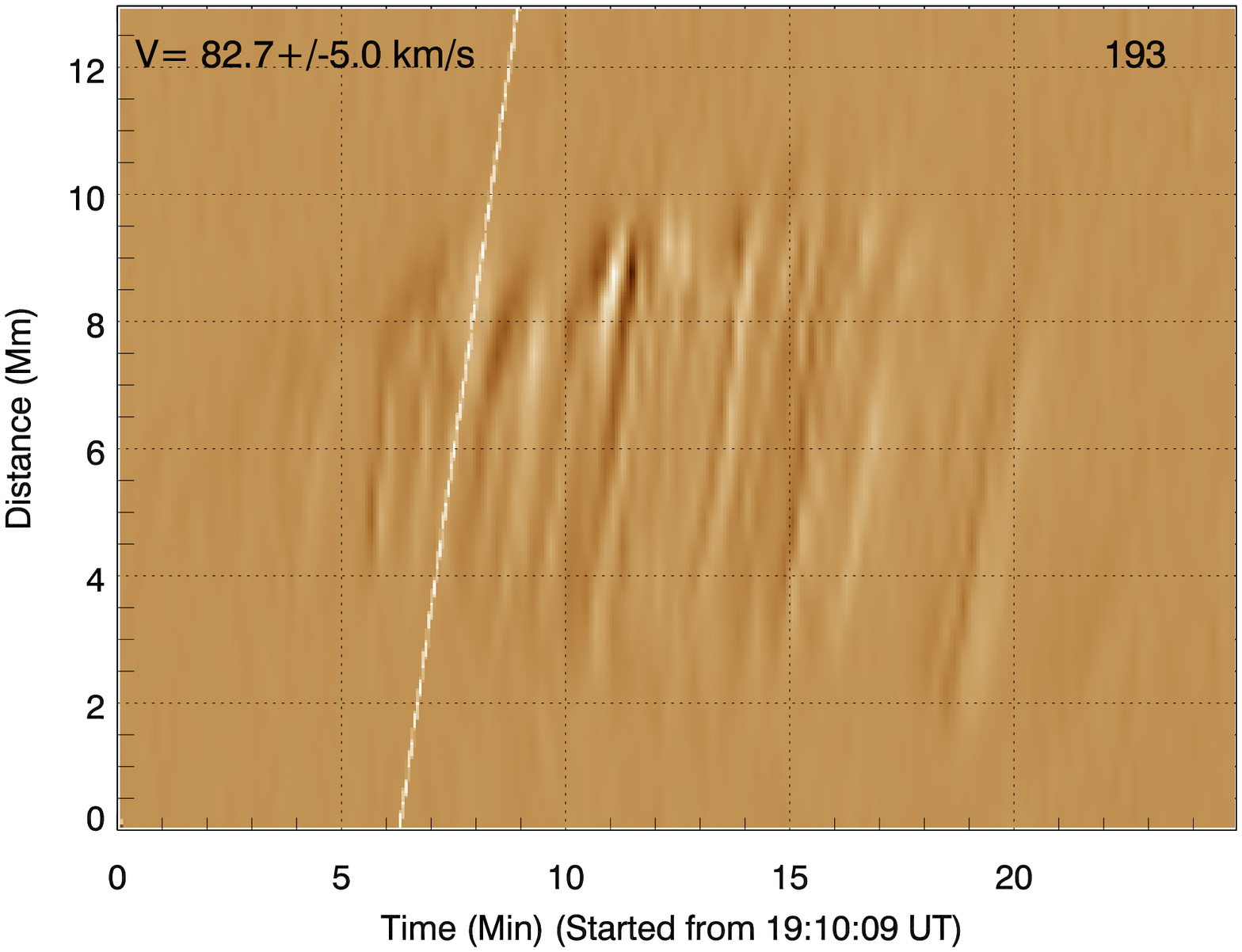}
}
\mbox{
\hspace{-1.0 cm}
\includegraphics[scale=0.4,height=18 cm,width=5.5 cm,angle=90]{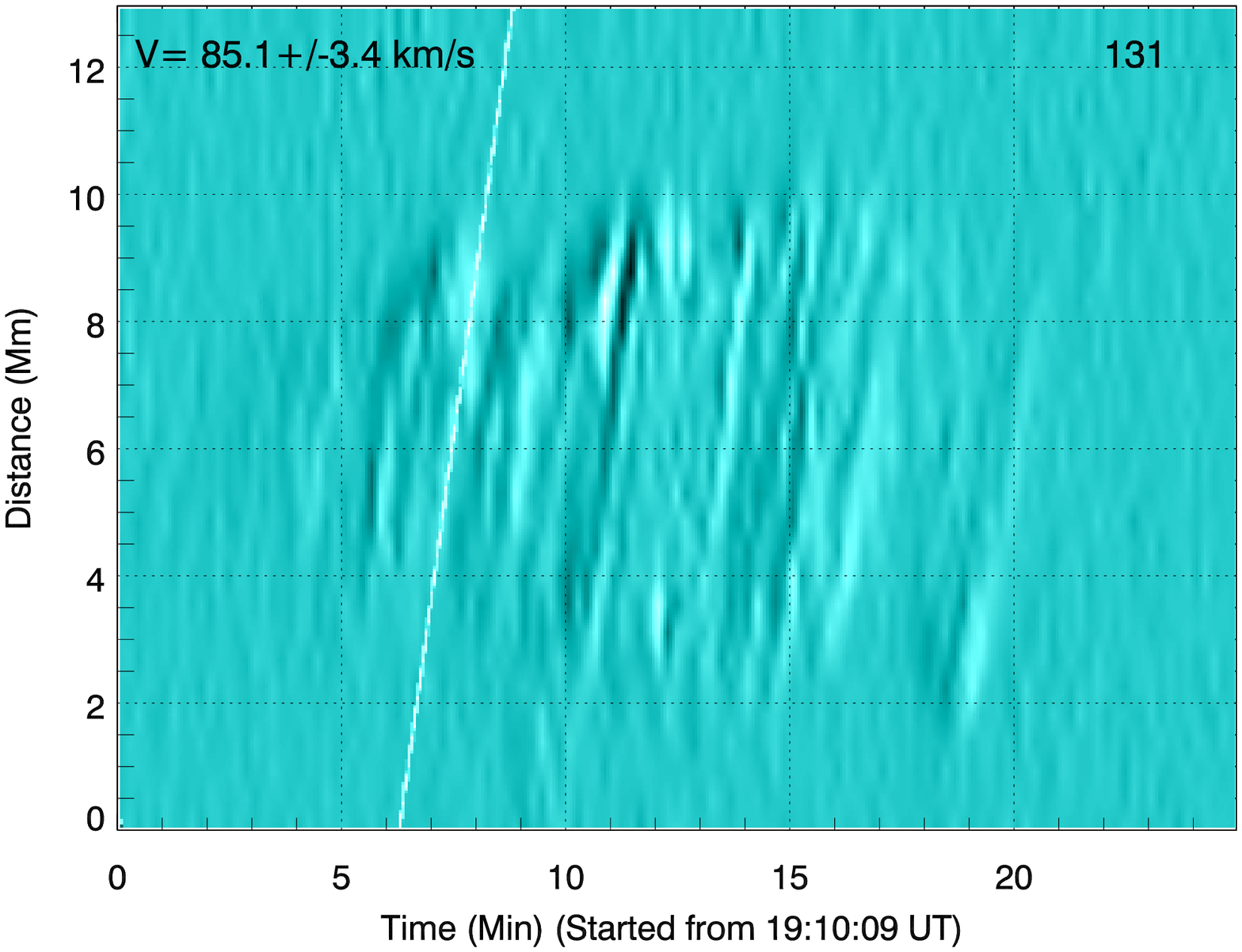}
}

\caption{Running difference distance-time maps at four different wavelengths, i.e., 304, 171, 193 and 131 {\AA} (successviely from top to bottom panel) estimated from slit shown in top-left panel of Fig. \ref{label 9}. Slopes shown as white straight line overplotted on these maps correspond to surfing speed necessary for the surfing signals meeting their resonance condition. Resolution along time step axis is 2.36 seconds and that along position axis is 0.09 Mm which are five times greater than original resolutions of AIA observations. This transformation has been performed by using REBIN function in IDL. The values of V basically suggest the surfing speed for which surfing signals reach their peak and therefore these values are considered as propagation speeds of PDs.} 

\label{label 10}
\end{figure*}

\section{RESULTS}

In the time span between 18:56 UT and 19:03 UT, non periodic faint plasma upflows having speeds 75 and 97 km \(\mathrm{s^{-1}}\) were observed to move towards the top of the cusp region along left magnetic segment (which is covered by a slit shown by green dashed lines from a to b in left-panels of Fig. \ref{label 3}) as evident in the distance-time map in right panels of Fig. \ref{label 3}. 

Just after those upflows, around 19:03 UT, two curved, elongated features indicative of magnetic flux-tubes were seen to approach each other as indicated by yellow arrows overplotted on the composite images of 171, 193 and 131 {\AA} in Fig. \ref{label 2} (see online animation related to Fig. \ref{label 2}). Thereafter an extended inverted Y-shaped magnetic geometry was found to be formed in the surroundings (the detailed analysis of magnetic topology is given in subsection 3.1.2). We term its upper part as an elongated cusp-shaped region. While, its downward part is the curved and closed loop-like structure. Around 19:09 UT, the elongated cusp-shaped portion of this magnetic structure was brightened which we termed as an apparent current sheet where reconnection might take place (see online animation associated with Fig. \ref{label 2}). Thereafter, additional brightness was seen to be originated at the top region of this elongated straight portion. The brightness propagated downward in the elongated cusp-shaped region and thereafter moved through the rightward magnetic channel. This generation of brightness and its propagation through that segment was observed almost continuously from 19:12 UT to 19:28 UT before the structure  disappeared (see online animation related to Fig. \ref{label 2}). \par 

To study the physical scenario associated with these propagating intensity features (termed as PDs hereafter) and nature of these features, we compute the DEM and magnetic structure in section 3.1, and we measure the period and propagation speeds at different wavelengths in section 3.2.

\subsection{Physical Scenario of the Region Associated with Propagating Disturbances (PDs)}

We conjecture that PDs were generated due to magnetic reconnection as two magnetic flux-tubes were clearly seen to approach towards each other and thereafter merge with each other just before the brightening was started. To examine whether magnetic reconnection is possible in this region, and therefore to provide conclusive support to our primary observational findings as shown in Figs. 1--2 about triggering of PDs, we tried to find out whether any heating was taking place in the considered region via differential emission measure analysis (see subsection 3.1.1). Also magnetic field extrapolation has been carried out for finding magnetic field topology in the ROI and thereafter that field topology was analyzed to locate magnetic nulls as well as QSLs as they are considered to be the preferential locations for an occurrence of magnetic reconnection (see subsection 3.1.2).

\subsubsection{Differential emission measure analysis and heating} \label{DEM}

We studied thermal structures of plasma flows via DEM analysis. We used the sparse inversion method as described by \citet{2015ApJ...807..143C} to estimate DEM at temperatures $\mathrm{\log}~T = 5.6-7.4$ using six AIA filters, i.e., 94, 131, 171, 193, 211 and 335 {\AA}. The total emission measure from the specified temperature interval, say, \({\Delta}T\) is 
\[\mathrm{EM_T}= \mathrm{DEM(T)} \mathrm{\Delta T} = \int_{0}^{\infty} \mathrm{n_e^2(T)} \mathrm{dl}\]
, quantifying the thermal characteristics of the plasma integrated over the portion of the line of sight (along the length element dl) per unit area on which the temperature is between T and T+\(\mathrm{\Delta T}\). \(\mathrm{n_e(T)}\) is the electron number density of the plasma at a certain temperature T \citep{2015ApJ...807..143C}. Since DEM is proportional to \(n_e^2\), it is considered as a proxy for studying evolution of number density of electrons. Also DEM weighted temperature was estimated in log scale using
\[\frac {\sum_{i=1}^{n} (DEM_{i}\times \mathrm{\log}~T_{i})}{\sum_{i=1}^{n} DEM_{i}}\]
, where \textit{i} stands for different temperature bins of width $\Delta \mathrm{\log}~T = 0.1$ and n is the total number of temperature bins. \par
We took a rectangular box (smaller box shown in DEM map for $\mathrm{\log}~T = 7.1-7.4$) of spatial extent x = [467\({\arcsec}\),472\({\arcsec}\)] and y = [-235\({\arcsec}\),-229\({\arcsec}\)] covering the top of the elongated cusp-shaped region from where the brightness was originated. The thermal nature of the considered event in different temperature ranges within $\mathrm{\log}~T = 5.6-7.4$ are shown in the top panel of Fig. \ref{label 4} (see online animation also). The temporal evolution of total emission measure (EM) as well as DEM weighted temperature are shown in the bottom panel of Fig. \ref{label 4}. It is evident that both of them show a rise starting around 19:12 UT and further peaks around 19:21 UT, and thereafter continue to decay. Since emission measure is proportional to the square of plasma number density, then an increase in emission measure may result from the increment of high temperature plasma within the ROI, i.e., in the elongated cusp-shaped region and in the downward magnetic channel (Fig. \ref{label 4}). \par

Now the instantaneous presence and increment of high temperature plasma can be due to either of two reasons or both, i.e., (i) some amount of plasma is already present there and gets heated by some mechanism (e.g., energy release due to reconnection), (ii) high temperature plasma is flowing into that region in due course of time, due to thermal conduction and evaporation of chromospheric plasma. We already observed faint plasma flows (prominently visible in cool AIA filter, i.e., 304 {\AA}) propagating upward along the left segment of the concerned magnetic structure to the ROI around 18:56-19:03 UT as discussed in Fig \ref{label 3}.

However, no flows were observed to enter in that region after that particular time during main epoch of the event (19:10 UT--19:28 UT). Therefore, the rise in the emission measure and temperature could not be caused by that particular upflow through left segment and rather resulted by heating of the plasma already present or filled there. \par

It is to be noted that there is a certain limitation of inversion methods used to estimate DEM using AIA filters alone that it does not  provide pinpoint accuracy in the estimation of EM in higher temperature bins for broader DEMs, i.e., for the multithermal plasma as AIA temperature response functions do not constrain higher temperature plasma accurately \citep{2012ApJS..203...26G,2015ApJ...807..143C}. However, in this work, our objective is to demonstrate the evolution of multi-temperature plasma contained within the considered structure during the magnetic reconnection process. We do not aim to estimate the exact amount of heating that occured in the evolved plasma to match it with the the excessive appearance of emitting plasma in high-temperature AIA filters. Instead, we only aim to show that there is an occurrence of some amount of heating to show that magnetic reconnection is an ongoing process that causes evolution of heated multi-temperature plasma flowing quasi-periodically in the concerned magnetic structure. Therefore, even if DEMs in high temperature are not well constrained, still it is clearly visible in higher temperature DEM maps that initially high temperature plasma was absent in the reconnection region and as reconnection sets in, they start to appear providing evidence of heating there (cf., Fig \ref{label 4}).
 
Therefore we infer that formation of X-type structure apparently seen in 2-D AIA images and merging of two flux-tubes just before the increments in both these quantities displays a connection between onset of magnetic reconnection and thereby possible heating and evolution of high-temperature plasma. \par   


\subsubsection{Measurements of magnetic field and associated magnetic nulls and QSLs}

In order to explore the occurrence of  magnetic reconnection with more conclusive evidence, we extrapolated magnetic field lines at coronal heights and tried to identify the presence of magnetic nulls/QSLs.
\newline
(I) \textit{Potential field extrapolation}--
\par 
The direct observation of magnetic field is only available at the photospheric level. Therefore, we used potential field extrapolation technique applying Green's function method \citep{1977ApJ...212..873C} to obtain magnetic field topology in the surroundings of our ROI located in the solar corona. The HMI \citep{2012SoPh..275..207S,2012SoPh..275..229S} vector magnetograms (\texttt{`hmi.B\_720s'} series) of temporal cadence 720 seconds were taken as our lower boundary conditions. Since the transverse component of those magnetograms were subjected to 180 degree ambiguity, magnetograms were corrected by removing this ambiguity using minimum energy method as implemented in the HMI pipeline \citep{1994SoPh..155..235M,2006SoPh..237..267M,2009SoPh..260...83L}. Since the ROI was far away from the disk center, the magnetograms were further corrected for projection effect by the method described in \citet{1990SoPh..126...21G}. Basically the magnetogram data taken from JSOC are in helioprojective cartesian coordinate system. So we deprojected it to the heliographic coordinate system. It was basically a two step procedure. Firstly, the components of vector magnetic fields in the helioprojective plane were transformed to heliographic components on the same helioprojective plane. Afterwards, the helioprojective plane has been transformed to a local plane tangent to the solar surface at the centre of the ROI. Since the geometry of field of view suffered change as a result of removing the projection effect, we recut the edges to get a rectangular boundary of the magnetogram enclosing ROI. The extrapolated area was resolved by 244 \(\times\) 132 grid points with coarsest spatial resolution in both $x$ and $y$ direction being 2\({\arcsec}\). Also, the force-free and torque-free conditions are usually not satisfied for the observed magnetic field and therefore a preprocessing method \citep{2006SoPh..233..215W} was taken into account to remove the net force and torque on the bottom boundary of the computational domain used for the extrapolation. The preprocessed magnetograms were used to extrapolate potential coronal magnetic field in MPI-Adaptive Mesh Refinement-Versatile Advection Code (MPI-AMRVAC)\footnote{\url{http://amrvac.org}} \citep{2012JCoPh.231..718K,2012ascl.soft08014V,2014ApJS..214....4P,2018ApJS..234...30X,2020arXiv200403275K}. Previous applications could be found in \citet{2013ApJ...779..157G}, \citet{2015ApJ...806..171Y} and \citet{2017ScChD..60.1408G}. 
The extrapolated field lines showed that there were multiple small scale loops at lower heights in the inner corona, and there were a few comparably elongated field lines which were originated from same polarities (negative) and had their other footpoints in a magnetic polarity far from the feature of interest as shown in Fig. \ref{label 5} (top panel).

(II) \textit{Magnetic nulls}-- 
\par
The extrapolated magnetic field topology was further analyzed using a range of methods (Poincar\'{e} index method and trilinear method) to compute the locations of magnetic nulls. But \citet{2020A&A...644A.150O} suggested that use of trilinear method is preferable for implementation to numerically simulated data with cartesian grid structure. So, we found presence of total 49 nulls in the entire computational domain of which 48 were in lower heights near the solar photosphere, while one is found in our ROI in solar corona applying trilinear method \citep{2007PhPl...14h2107H,2020A&A...644A.150O} in Null\(\_\)Finder\footnote{\url{https://github.com/FedericaChiti/Null_Finder/tree/v1.0.0}} module. The exact position of this null is found to be x = 468 \({\arcsec}\) and y = -240 \({\arcsec}\). Therefore, we confirm the presence of one magnetic null within ROI and hence provide evidence for preferential topological feature associated with onset of magnetic reconnection.

(III) \textit{Quasi separatrix layers (QSLs)}--
\par
Also, we carried out an estimation for finding presence of QSLs within the same domain \citep{2013ApJ...779..157G,2015ApJ...806..171Y} using magnetic modeling codes\footnote{\url{https://github.com/njuguoyang/magnetic_modeling_codes}} module. We calculated the squashing factor (Q) which is a function defined by \citet{2002JGRA..107.1164T} to characterize QSLs. Basically, measurement of Q will determine the aspect ratio of the distorted ellipse exhibited due to mapping of an elementary flux-tube having its footpoint of positive (negative) polarity in an infinitesimal circular region to the footpoint of negative (positive) polarity sign i.e., Q quantifies how much squashing is associated with the initial elementary region as a result of field-line mapping from one to other footpoint \citep{2009ApJ...705..926B}. 

Mathematically, QSLs are defined by Q \(\gg\) 2 \citep{2002JGRA..107.1164T,2009ApJ...705..926B}. In our case, we found that Q had a maximum value as high as \(10^{10}\) and the path traced by prominent positions of QSLs resembled spine fan topology (Fig. \ref{label 5}; Bottom panel) with a magnetic null at the intersection of the spine axis and the fan plane as evident in Fig. \ref{label 6}. So in 3D volume of the observed region, magnetic null is also present apart from enhanced Q value. Therefore we consider that this region with QSLs along with presence of magnetic null, i.e., SLs is favourably causing periodic reconnection \citep{2017ApJ...844....2T,2019A&A...621A.106T}.\par 
 
The elongated cusp-shaped region of QSLs or SLs resemble with the same feature evident in the EUV intensity (Figs \ref{label 1}-\ref{label 2}) and DEM (Fig. \ref{label 4}). The pre-existing  non periodic plasma upflows (Fig. \ref{label 3}) during 18:56-19:03 UT transported mass from left side of such magnetic field configuration to such QSL (or SL) region on its top. The QSL region containing a magnetic null is the place where multiple reconnection took place after 19:10 UT when heating occurred and intensity disturbances began to propagate towards the downward magnetic channel around 19:15 UT (see Fig. \ref{label 7} and related animation). So presence of heating at the top of elongated cusp-shaped region, QSL or SL resembling the path followed by the strong plasma flows along with presence of magnetic null at the intersection of spine axis and fan plane are providing conclusive evidence of occurrence of magnetic reconnection in this ROI. Therefore magnetic reconnection is established as driving mechanism of the observed heating of the plasma and generation of PDs.

\subsection{Nature of propagating disturbances}

Successive bright and dark fronts were observed to propagate downward along the cusp-shaped region (which resembles now with QSL with a null in its lower segment) in the running difference images of 304, 171, 193 and 131 {\AA} (from top to bottom panel) in Fig. \ref{label 7} (see related animation also). Presence and propagation of these successive bright and dark fronts along field lines often suggest to consider PDs as slow magnetoacoustic waves. But it is not straightforward as shown in some literature as well as in this paper. Now the methods used for analysis of the observed PDs for searching  propagation speeds and periodicity are described here in a sequential manner along with their corresponding results. We used the method described in \citet{2012A&A...543A...9Y} to extract the distance-time maps. We took a straight slit S1 having width of 5 pixel and length of 13.20 Mm as shown in top-left panel of Fig. \ref{label 9} along the straight part of the considered magnetic geometry along which propagating disturbances were noticed to propagate. Now it is clear that this elongated cusp-shaped region resembles the QSL in shape (bottom panel; Fig. \ref{label 5}), and the PDs actually propagated in it downward. We estimated distance-time maps for normal intensity images. The distance time diagrams for four wavelengths 304, 171, 193 and 131 {\AA} are shown in top right panel of Fig. \ref{label 9}. The bright ridges were seen in distance-time maps, which represent the propagation of PDs with time (as evident from tilted representation of those ridges) in all of these wavelengths. \par

Although we analyzed four AIA bandpasses (304, 171, 193 and 131 {\AA}) only, the PDs were visible in all SDO/AIA bandpasses except 94 and 335 {\AA} as well as in different temperature ranges in DEM (as shown in animation related to Fig. \ref{label 4}). Therefore we infer that the PDs are multi temperature in nature. We used Si IV 1393.77 {\AA} optically thin lines (since the corresponding emission can be considered as cooler counterpart of the EUV emissions observed via SDO/AIA, and almost similar to the formation temperature of AIA He II 304) to spectroscopically constrain the reconnection region. 

We have taken Si IV line spectra from a few pixels around the red cross symbol (as shown in left panel of Fig. \ref{label 8}) from each raster file. Basically raster scan is repeated at red crossed region at time interval of 40 seconds and we get spectral line profile of Si IV over the reconnection region. Further, the averaged spectral profile is fitted by a single Gaussian to estimate line width variation in time. Then we found that line width is subjected to gradual rise (red-line in right-panel of Fig. \ref{label 8}) followed by peak around 19:21 UT and sharp decay after that (Fig. \ref{label 8}; right panel). This time-line is identical to the rise and fall of EM and Temperature in the ROI, and overall the occurrence of the considered event. This line width increment is associated with the evolution of non thermal velocity (unresolved otherwise) of mass motion due to underlying turbulence \citep{1977ApJ...212L.143D}, quasi-periodic upflows or waves or shocks \citep{2012ApJ...759..144T,2015ApJ...799L..12D}. So the line width increment reconfirms the presence of plasma flows generated at the top of the elongated cusp-shaped region and thereafter propagated downwards along the straight part of the cusp region. 
 
\subsubsection{Periodicity in observed PDs}

We extracted light curves exhibiting the evolution of intensity oscillations, for example at a height of 6 Mm as denoted by horizontal magenta lines overplotted on distance time maps shown in Fig. \ref{label 9}. The corresponding normalized light curves for four wavelengths, i.e., 304, 171, 193 and 131 {\AA} obtained by dividing the data points by their maximum values are shown in the top sub-panel in the bottom panel of Fig. \ref{label 9}. We performed wavelet analysis using Morlet wavelet as basis function to estimate the periodicity of the oscillations \citep{1998BAMS...79...61T}. The wavelet power spectrums for different wavelength filters are shown in the bottommost sub-panel in the bottom panel of the Fig. \ref{label 9}. The cross-hatched area on each wavelet power spectrum outlines the cone of influence (COI) where power is not considered reliable due to edge effects. We have computed 95 {\%} local significance levels using the power law model introduced by \citet{2016ApJ...825..110A} and further used by \citet{2020A&A...634A..63K} and \citet{2022MNRAS.517..458S}. If \(\nu\) is the fourier frequency, power law equation (with A and C being constants) is given as :
\begin{equation}
\sigma(\nu) = A\nu^{s}+C
\end{equation}
We have fitted the Fast Fourier Transformation (FFT) of each light curve with this power law equation (as given above). Using the fitted curve as the background spectrum, we have estimated the 95 {\%} local significance level \citep{2016ApJ...825..110A}.\par

In the sub-panel between light curves and wavelet power spectrums, the FFTs of the light curves are shown in black color, the global wavelet spectrums (time-averaged of the wavelet power spectrum) are shown in red color. The fitted curves using the power law equation are shown in blue color, 95 {\%} local significance levels are shown in the green color. If the global wavelet power is above these 95 {\%} local significance levels, only then it is considered significant \citep{1998BAMS...79...61T}. We have excluded the power lies within COI and overplotted the 95 {\%} local significance contour on the wavelet power spectrum. Wavelet spectra have significant power within the range of 2-6 minutes, and the power is dominant at 4.28 minutes in all wave bands, i.e., 304, 171, 193 and 131 {\AA}, as can be seen from the global power spectrum. This clearly elucidated the quasi-periodic nature of the PDs. To infer that whether these quasi-periodic PDs are waves or flows, we estimated their speeds in four different wavelengths, i.e., 304, 171, 193 and 131 {\AA} as described in subsection 3.2.2.  \par

\subsubsection{Propagation speeds at different wavelengths}

We used the surfing-transform technique \citep{2013ApJ...778...26U,2021ApJ...907....1U} to estimate the average speeds of quasi-periodic PDs at four different wavelengths, i.e., 304, 171, 193 and 131 {\AA}. We calculated the surfing transform of the running difference distance-time maps derived by temporal difference of 12 seconds (as shown in Fig. \ref{label 10}) for surfing speeds within range of 20 to 250 km \(\mathrm{s^{-1}}\). We took the step size for increment of surfing speeds during measurement to be 0.5 km \(\mathrm{s^{-1}}\). We inferred those surfing speeds as average phase speed which yielded strongest surfing signal as the signal met resonance condition (for more details, see \citealt{2013ApJ...778...26U}). In those used maps shown in Fig. \ref{label 10}, the resolution along horizontal time axis is 2.36 seconds and that along vertical position axis is 0.09 Mm. Actually the resolution on both the axes, i.e.,  time and distance were increased by 5 times to that of AIA observations using the REBIN function in IDL which gives an interpolation of the original maps. Basically, one pixel is subdivided in 5 subpixels for better visualization. Now in surfing-transform technique, during the error estimation, there are two major error sources: (a) the uncertainty in the peak determination and (b) the uncertainty caused by the signal nonstationarity. The former depends on the length of the slit, the frequency of the oscillation, and the peak location (for more details, see \citealt{2013ApJ...778...26U}). The peak uncertainty (a) in our data is relatively small, of about 1-2 km \(\mathrm{s^{-1}}\). The nonstationarity error (b) is substantially higher and considered as the main factor limiting the accuracy of the results. To estimate the nonstationarity error, we used the boxcar method. For each of four AIA channels, we computed the local surfing velocities for a set of 12.5 minute time intervals with a varying starting time, to evaluate the variability of the entire 25 minute time interval. It is to be noted that these local estimates are not statistically independent. But this is expected as we are dealing with an uncertainty caused by nonstationarity as the latter results from a combination of physical processes manifesting themselves as a long-term trend, with a significant autocorrelation time. It is this low-frequency process that makes the most significant contribution to the uncertainty. We used overlapping boxcar windows due to the limited duration of studied solar events, but even if they were long enough to apply non-overlapping windows, the local estimates would not be statistically independent because of the physical origin of the nonstationarity. We found that the Boxcar-averaged corrected surfing speeds corresponding to the peaks of surfing signals i.e., resonance of surfing signals as estimated accurately using parabolic fitting of surfing RMS vs surfing speed profile \citep{2013ApJ...778...26U} were $94.7\pm 4.0$ km \(\mathrm{s^{-1}}\) for 304 {\AA} passband, $90.1\pm2.5$ km \(\mathrm{s^{-1}}\) for 171 {\AA} passband, $82.7\pm5.0$ km \(\mathrm{s^{-1}}\) for 193 {\AA} passband and $85.1\pm3.4$ km \(\mathrm{s^{-1}}\) for 131 {\AA} passband. It is to be noted that the above mentioned surfing RMS is the  root mean square of the pixel intensity. Its physical units are the same as the ones used in the processed AIA FITS files. \par

Now the coronal plasma is actually multi-thermal in nature and the emission at various temperature is narrow band emission. Also, emission at 304 {\AA} wavelength is associated with an even more cooler temperature (log T = 4.7) than other three wavelength filters. So, keeping these in mind, it can be inferred that the plasma dynamics under consideration is truly multi-thermal in nature. In this circumstance, the estimated speeds at different wavelengths, i.e., at different temperatures clearly suggest that these speeds are not scaled as \(\sqrt{T}\) as expected for slow magnetoacoustic waves, where T stands for local temperature \citep{2004psci.book.....A}. For example, the ratio of theoretically predicted speeds at 304 {\AA} and 171 {\AA} is 0.27 but that for estimated speeds at 304 {\AA} and 171 {\AA} is within $0.98-1.13$. If we consider the estimated speed at 304 {\AA} as our reference speed and try to find the theoretically expected speeds, the speeds at other filters will be higher than that in 304 {\AA} which is not the case here. In general, the speeds are not increasing with the increment in the peak temperatures across different AIA wavebands and therefore are not scaled with \(\sqrt{T}\). In addition to this, since we are using imaging observations which depend upon density perturbations to probe the presence of any disturbances, presence of {Alfv\'en} waves can be safely discarded as it can not be probed by perturbation in density because of its incompressible nature \citep{2014masu.book.....P}. Also, for fast waves, the propagation speeds are expected to be much larger (roughly of an order of 1000 km \(\mathrm{s^{-1}}\) in the solar corona) than these speeds estimated using Surfing technique \citep{2011ApJ...736L..13L,2012ApJ...753...52L,2014SoPh..289.3233L,2017ApJ...851...41Q}. 
 Therefore the PDs are essentially the flows but certainly not the slow waves with phase speed scaled as \(\sqrt{T}\) or any other form of waves like fast magnetoacoustic waves or {Alfv\'en} waves.\par  
 
\section{DISCUSSION \& CONCLUSION} 
Using SDO/AIA, we observed two narrow elongated curved features (magnetic flux-tubes) to approach towards each other around 19:03 UT and thereafter merge with each other after a few moments. Prior to this merging, the non periodic plasma upflows from left side of this region (as described in Fig \ref{label 3}) filled those particular strands that were merging. The thermal and magnetic measurements are consistent with magnetic reconnection in this elongated cusp-shaped region, which is essentially the QSL region more prone to the magnetic reconnection. The spectroscopic measurement of Si IV line width increment is also consistent with the subsequent mass motion during the ongoing reconnection there. Intensity disturbances started to generate as a result of the inferred magnetic reconnection and propagated downward along the cusp part formed by merged field lines and thereafter predominantly along extended right segment of field lines just after the merging. \par

As mentioned in above summary, DEM analysis provided us signature of occurrence of heating in the top of the QSL region. This is linked with the signature of the occurrence of magnetic reconnection. The magnetic topology of the extrapolated field lines resembled closely the plasma emission structures as seen in SDO/AIA images. QSLs were found to be present in the ROI resembling a spine and fan topology having a magnetic null at the intersection of the spine axis and the fan plane. The presence of magnetic null in the extrapolated field provides strong supporting evidence for onset of magnetic reconnection in the ROI on the basis of favourable magnetic topology. 
Along with the magnetic null, QSLs are also preferential positions for occurrence of magnetic reconnection as those are the regions where magnetic field lines possess drastic changes in their connectivity even though they are not purely discontinuous, i.e., oppositely directed to each other \citep{1994ApJ...437..851L,1995JGR...10023443P,1997A&A...325..305D,1997SoPh..174..229M,1999ApJ...521..889M,2005A&A...444..961A,2008ApJ...675.1614T}. \par

The PDs show periodicity of about four minutes as estimated via wavelet analysis. As shown in Fig. \ref{label 7}, the PDs were propagating as successive bright and dark fronts in running difference images which initially were suggestive of slow magnetoacoustic waves. However, this interpretation is not supported by the surfing transform analysis. If  \(\phi\) is an inclination angle of the coronal field lines along which the intensity disturbances propagate with the line of sight, then the projected propagation speed of the disturbances, say, \(V_p\) will be equal to \(V_{s} \mathrm{\sin}~\phi\) where \(V_{s}\) is the actual speed \citep{2009A&A...503L..25W}. As the spatial region under consideration was far from centre of the disk, projection effect came into picture. Therefore, depending upon the inclination angle \(\phi\), the projected propagation speed which we were estimating would be smaller than actual speed. Nevertheless, if the disturbances were really slow mode magnetoacoustic waves, the propagation speeds, even if they would be smaller than local sound speed, say, \(C_{s}\), would follow \(\sqrt{T}\) dependence as discussed in \citet{2004psci.book.....A}. But the measurement of propagation speed of PDs at different wavelengths, i.e., at different temperatures exhibited the speeds which did not show \(\sqrt{T}\) dependence. Also, the presence of other forms of waves like fast magnetoacoustic waves or {Alfv\'en} waves are discarded as explained in section 3.2.2. Therefore, we describe these observed propagating disturbances as quasi- periodic plasma flows instead of the slow magnetoacoustic waves. \par

Various driving mechanisms were proposed recently to be associated with ubiquitous plasma flows since their discovery via  Hinode/EIS observations \citep{2008A&A...481L..49D,2009ApJ...694.1256T,2012ApJ...754L...4T,2014SoPh..289.4501S}. Chromopsheric evaporation as a result of reconnection due to flux emergence and braiding of field lines by random motions at photospheric level, footpoints of active region loops subjected to an impulsive heating, evolution of large-scale reconnecting loops etc, were carried forward as driving mechanisms of those outflows \citep{2008A&A...481L..49D,2008ApJ...678L..67H,2008ApJ...676L.147H}. One possible mechanism behind generation of periodic reconnection outflows at a 3D magnetic null is oscillatory reconnection \citep{2017ApJ...844....2T}, although the expected periodicity of this mechanism in the corona is not yet well constrained \citep{2019A&A...621A.106T}. Along with an existence of null point resulting in magnetic reconnection,  generation of plasma flows in ARs are also related to the presence of QSLs and occurrence of reconnection there as firstly proposed by \citet{2009ApJ...705..926B}. Using Hinode/EIS, \citet{2012SoPh..281..237V} observed AR related plasma outflows and measured their speeds. They performed magnetic-field extrapolations of AR and and confirmed co-spatiality of outflows with QSL locations, including the separatrix of a magnetic null. \citet{2013SoPh..283..341D} found that the coronal plasma upflows from the edges of AR 10978 were thin, fan-like structures rooted in QSLs between high pressure AR loops and less pressure loops in the neighbourhood. Also \citet{2015ApJ...809...73M} reported that AR plasma upflows observed by EIS and QSLs located by analyzing extrapolated field lines were found to evolve in parallel, both temporarily and spatially. \citet{2017SoPh..292...46B} and \citet{2021SoPh..296..103B} had carried out similar analysis and interpretation of driver of plasma upflows in ARs. Hence it has been inferred that strong and dominant QSLs (with or without magnetic nulls) are the preferential positions for accumulation of strong current and occurrence of magnetic reconnection resulting in observed plasma outflows in solar ARs.

In this paper, we analyzed multiwavelength imaging observations of quasi-periodic plasma flows and found presence of magnetic null as well as dominant QSLs resembling the path followed by the strong outflows. We observed the flow to be asymmetrically propagating as it was mostly propagating along the right segment of an inverted Y-shaped magnetic channel as seen in the plane of the sky (2D) AIA images. We conjecture that the presence of field lines having larger strength was leading to dominant movement of plasma flows in that direction as compared to the relatively weaker field lines in the left segment. In this work, in addition to locating presence of magnetic null in the ROI and matching position of QSLs with strong plasma flows, we examined heating in the top of the elongated cusp-shaped region, where the bundle of field lines or flux-tubes were merging with each other, via differential emission measure (DEM) analysis and therefore provided a detailed evidence of magnetic reconnection. Also since we implemented surfing transform method to compare the propagation speeds at different wavelengths or different temperatures, it came out that the speeds at four different wavelengths, i.e., at different temperatures were not scaled as \(\sqrt{T}\) and therefore provided confirmation about our interpretation of the plasma flows instead of slow magnetoacoustic waves. Also, we performed wavelet analysis to examine the periodicities in the PDs. We found quasi-periodicity of about four minutes associated with these PDs. Therefore we infer that the reconnection which was acting as driver of these quasi-periodic plasma flows may be taking place periodically in QSL or SL region (akin of elongated cusp shaped region as seen in AIA images). The bundle of field lines or flux-tubes might be undergoing repetitive merging due to changes in pressure gradient between them. The faint plasma upflows (discussed in Fig. \ref{label 3}) which were observed to propagate through various channels almost repetitively might have established some additional plasma pressure and therefore total pressure which may lead to rightward movement of left field line bundles resulting in repetitive merging of it with the bundle of field lines on the right side and hence in periodic reconnection. Also, the in-phase fluctuations of EM and DEM weighted temperature (as shown in bottom panel of Fig. \ref{label 4}) may be suggesting that quasi-periodic flows are produced by intermittent heating during the reconnection process. While the detailed physical picture we present in this paper regarding magnetic reconnection generated quasi-periodic plasma flows in the observed event, we expect that additional high-resolution and high cadence imaging observations in the future will confirm such evolutions at many different QSLs/SLs evolving in various parts of solar corona. \par

In previous literature, in general, the quasi-periodic intensity (thus density) fluctuations in the distance-time maps as well as light-curves were demonstrated as the signature of magnetoacoustic waves (e.g., \citealt{1997ESASP.404..571O,2000A&A...355L..23D,2009A&A...499L..29B,2009ApJ...697.1674M,2011A&A...528L...4K,2011A&A...526A..58S,2012A&A...546A..93G,2013ApJ...779L...7K,2015ApJ...804....4K,2017A&A...600A..37N}). We are not in disagreement with those physical scenarios of magnetoacoustic waves and they must remain valid in such cases of plasma dynamics seen in the localized solar atmosphere. However, in this paper, using theoretical argument of \(\sqrt{T}\) dependence of propagation speed along with application of Surfing technique for stringent analyses of speeds, we find that if \(\sqrt{T}\) dependence of phase speeds does not hold true for any specific physical dynamics and the estimated phase speeds are much less than the speeds expected for fast magnetoacoustic waves, the presence of quasi-periodic intensity fluctuations or successive bright and dark tilted ridges in distance-time maps can be a signature of quasi-periodic flows instead of magnetoacoustic waves. So, through this work, we put forward the fact that there will not always be a one to one correspondence between the presence of successive bright or dark ridges in distance-time map and signature of magnetoacoustic waves \citep{2012ApJ...754...43S,2014ApJ...793...86P}. We suggest that similar observables can be linked with different physical processes at any particular epoch of time in the localized solar atmosphere, thus caution should be taken in the form of stringent analysis before making physical interpretations. Although the waves and flows can also co-exist in many cases, here our detailed analyses demonstrate alone the origin of quasi-periodic plasma flows due to reconnection in QSL/SL above the magnetic null in the localized corona.

\begin{acknowledgments}
We are thankful to the anonymous reviewer for his/her valuable suggestions and comments which help us to improve this manuscript. S.M. thanks Mark Cheung for his valuable comments and discussions regarding DEM analysis. Y.G. was supported by NSFC (11773016 and 11961131002) and 2020YFC2201201. DP gratefully acknowledges support through an Australian Research Council Discovery Project (DP210100709). TJW and LO acknowledge support by NASA grants 80NSSC18K1131, 80NSSC21K1687 and 80NSSC22K0755. LO, TJW and VMU acknowledge support by NASA Partnership for Heliophysics and Space Environment Research (PHaSER) award 80NSSC21M0180. A. K. Srivastava acknowledges the ISRO Project Grant (DS\_2B512 13012(2)/26/2022-Sec.2) for the support of his research.
D.Y. is supported by NSFC 12173012, 12111530078 and the Shenzhen Technology project (GXWD20201230155427003-20200804151658001)
\end{acknowledgments}

\vspace{5mm}
\facilities{SDO/AIA, SDO/HMI, IRIS}
\software{MPI-AMRVAC, SolarSoft}




\end{document}